\documentclass[american,aps,pra,reprint,superscriptaddress,onecolumn]{revtex4-1}
\usepackage{amsmath,amscd,amsthm,amssymb}
\usepackage{mathrsfs}
\usepackage{graphicx}
\usepackage{epsf,epstopdf}
\usepackage[all]{xy}
\usepackage[unicode=true,pdfusetitle, bookmarks=true,bookmarksnumbered=false,bookmarksopen=false, breaklinks=false,pdfborder={0 0 0},backref=false,colorlinks=false] {hyperref}
\hypersetup{colorlinks=true,linkcolor=myurlcolor,citecolor=myurlcolor,urlcolor=myurlcolor}
\usepackage{graphics,epstopdf,graphicx, amsthm, amsmath, amssymb, times, braket, color, bm}
\usepackage{xcolor}
\usepackage{cleveref}
\usepackage{caption}
\usepackage{subcaption}
\usepackage{makecell}

\definecolor{myurlcolor}{rgb}{0,0,0.7}

\theoremstyle{plain}

\providecommand{\theoremname}{Theorem}
\newcommand*{\myproofname}{Proof}

\makeatother

\theoremstyle{definition}

\theoremstyle{remark}

\begin{document}

\title{Quantumly controlled measurement, Hermitian conjugation and normalization in matrix-manipulation algorithms}


\author{Edward B. Fel'dman}
\email{efeldman@icp.ac.ru}
\affiliation{Federal Research Center of Problems of Chemical Physics and Medicinal Chemistry RAS,
Chernogolovka, Moscow reg., 142432, Russia}

\author{Alexander I. Zenchuk}
\email{zenchuk@itp.ac.ru}
\affiliation{Federal Research Center of Problems of Chemical Physics and Medicinal Chemistry RAS,
Chernogolovka, Moscow reg., 142432, Russia}

\author{Wentao Qi}
\email{qiwt5@mail2.sysu.edu.cn}
\affiliation{Institute of Quantum Computing and Computer Theory, School of
Computer Science and Engineering,
Sun Yat-sen University, Guangzhou 510006, China}


\author{Junde Wu}
\email{wjd@zju.edu.cn}
\affiliation{School of Mathematical Sciences, Zhejiang University, Hangzhou 310027, China}


\begin{abstract}

\noindent In this paper, we  solve three important problems  that are
revealed, in particular, to matrix-manipulation algorithms.
The principal  novelty is introducing the concept of  quantumly controlled
measurement  that removes the post-selection problem by solving the problem of small access probability  to the
desired state of ancilla and possesses several  remarkable properties. We
also introduce  separate  encoding of the  real and imaginary parts of a
complex matrix that allows to include  the Hermitian conjugation into the
list of matrix manipulations.
Finally, we weaken the constraints on the { modulus} of  matrix
elements unavoidably imposed by the normalization condition for a pure
quantum state.  The quantumly controlled measurement together with both  other
extensions are implemented into the matrix multiplication algorithm. The
appropriate circuits are presented.

\end{abstract}

\maketitle

\newpage

\section{Introduction}
Quantum algorithms represent  a rapidly developing area of quantum informatics whose progress is stimulated by prospective enormous advantages of quantum algorithms over their classical counterparts. We have to acknowledge such famous algorithms as
 the Deutsch algorithm~\cite{Deu} demonstrating effect of  quantum parallelism in operation with Boolean functions, Shor's factoring algorithm~\cite{Sho1, Sho2}  utilising  the quantum Fourier transform~\cite{QFT1, QFT2, QFT3},  Grover's search  algorithm~\cite{Gro}, phase-estimation algorithm for the Hermitian \cite{QFT3} and  non-Hermitian \cite{Wang}  matrices .

We consider a particular application area of quantum algorithms which covers matrix algebraic  operations, such as multiplication, addition, determinant calculation, matrix inversion and its application to linear system solver.
During recent years, there are many quantum algorithms  solving various aspects of linear algebra. Among them  are HHL-algorithm for solving systems of linear equations \cite{HHL,HHL1,HHL2,HHL3,HHL4,HHL5,HHL6,HHL7} which was applied, in particular,  to Gaussian process regression \cite{ZhaoZ},  algorithm for solving linear differential equations \cite{Berry}, algorithm for evaluation of  matrix functions \cite{Tak}.

A possible approach
 to quantum algorithms for matrix manipulations is proposed in \cite{ZZRF}. This approach refers to matrices as operators to be applied to some vector. Therefore, it is essential in that approach that arbitrary nonunitary and non-Hermitian matrix  must be  transferred to the unitary one { before been included into the quantum algorithm}.  Thus, first of all, one has to replace  a given matrix $A$ (rectangular in general) with the Hermitian one
and then
  transfer it  to the unitary operator by exponentiating
which can be approximated via Trotterization \cite{Trotter,Suzuki,T,AT} and Baker-Champbell-Hausdorff  method \cite{Feldman,BC}. To implement such matrix transformations effectively,   the set of special matrices is proposed to be  prepared in advance. The phase estimation subroutine \cite{QFT3} is included into that algorithm.

However, mentioned algorithms do not cover all achievement  in matrix-manipulation techniques. A special block-encoding model \cite{GSLW}  is another approach for implementing matrix algebra.
Different type of block-encoding  was used in the algorithm in Ref.\cite{DFZ_2020} to embed the inverse of the
matrix  into the unitary transformation.
{In \cite{LWWZ},  matrix multiplication was performed  trough the binary encoding of matrix elements into the pure states of quantum systems  with subsequent binary multiplication via Toffoli gates. Matrix multiplication over  some rings, in particular,  multiplication of Boolean matrices, was studied in \cite{KN}.}

Some algorithms for matrix inversion have been also developed. As the first one we call the mentioned above HHL algorithm \cite{HHL}  where the  inversion  was used for solving  a  linear system. However, the problem of inversion of matrix eigenvalues has not been resolved in frames of  quantum algorithm  therein.
The  matrix inversion  algorithm based on the singular value decomposition   is presented in \cite{MRTC, NJ}. In both papers the authors apply  the function evaluation problem \cite{GSLW}  to approximate the inverse of singular values by odd-polynomial. This approach  requires introducing a special scale parameter  depending on the interval where all nonzero  singular values are  distributed.

Recently the alternative  matrix-manipulation algorithms based on the  encoding the matrix elements into the complex probability amplitudes of a pure superposition state of a certain subsystem  have been developed \cite{QZKW_arxive2022,ZQKW_2024,ZBQKW_arXive2024}. The principal difference of such approach is that a matrix is considered as a part of the quantum state of a system rather than  a unitary operator applied to this system.  Therefore the  elements  of the resulting matrix (obtained after addition or  multiplication of other matrices, or after matrix inversion) can be measured up to some constant factor.
The algorithms for matrix manipulations, such as inner product, sum and product of two  matrices, calculation of determinant, matrix inversion, and solving linear systems, based on special unitary operators applied to the quantum system were proposed in  \cite{QZKW_arxive2022}.
The realization of those unitary operators in terms of the basic quantum operations was explored later
 in \cite{ZQKW_2024} (inner product, matrix addition and multiplication) and then in \cite{ZBQKW_arXive2024}
 (calculating determinant and inverse matrix, linear system solver). Those unitary operators are based on the  multi-qubit controlled operators and are supplemented by the one-qubit  ancilla measurement at the final stage of the algorithm   for selecting the useful result and removing the garbage.
The major disadvantage of those  algorithms is the small probability of obtaining the required ancilla state after measurement (small success probability). This probability decreases polynomially (addition, multiplication) or exponentially (determinant calculation, matrix inversion) with the dimension of the considered matrix thus reducing efficiency of  the matrix-manipulation algorithms.

 The principal novelty of our paper is that we replace the above measurement of the one-qubit ancilla state with the  so-called quantumly controlled measurement (QCM), i.e., the measurement controlled by the quantum state of another one-qubit ancilla. In other words, we replace the one-qubit ancilla with the two-qubit one so that the state of the first qubit governs the measurement operator applied to the second qubit. This formally simple but crucial modification removes the above disadvantage of small success probability.
 { Unlike the algorithm implementing the usual measurement, which   requires   multiple running  to  reach the desired ancilla state,  the algorithm  involving   the QCM   assumes  the single run to
 achieve the goal.} This is the main advantage of the new algorithm.
  Such QCM can be included into all the algorithms in Refs.\cite{ZQKW_2024,ZBQKW_arXive2024}  as a subroutine replacing the usual ancilla measurement and thus significantly increasing their relevance. We note that the
QCM  has been implemented in Ref. \cite{ZBQKW_arXive2024} without {  full description.} In our paper, we  present the detailed analysis of this concept. The phenomenon  of quantum control  of  classical operations (measurement), i.e., the control  by the quantum state of the ancilla  qubit,  is a consequence of  the deep relation between the quantum and classical physics and must exist  relying on reality of quantum superposition states. {The demonstrated advantage of the QCM over the usual  measurement motivates research on its particular realization in terms of  acknowledged quantum and classical operators.}
{ The applicability of QCM can be extended to all measurement based quantum algorithms where the success probability is of principal meaning.  This refers to the measurement-based quantum algorithms \cite{WWLN,ZDB,ZBD,Wei,WeiBook}, in particular, to  the HHL-algorithm for solving linear systems \cite{HHL,HHL1,HHL2,HHL3,HHL4,HHL5,HHL6,HHL7};   various applications of measurement in quantum algorithms are reviewed in \cite{USBGSN}. }

Along with implementing the QCM, we also introduce two following  extensions to the matrix-manipulation algorithms.
(i) Although those  algorithms deal with complex matrices in general, some operations (including Hermitian conjugation) can not be realized in frames of those settings and therefore certain constraints  are to be imposed on the variety of matrix manipulations. To remove this disadvantage  and construct  more flexible  algorithms for matrix algebra  we propose to separate  the real and imaginary parts of a matrix {  by encoding them  into two orthogonal subspaces of the whole state space.} This separation can be done by involving additional one-qubit subsystem  whose state  serves to label  the  real and imaginary parts of  matrix elements.  As a consequence, all probability amplitudes are real in such representation. We notice that the Hermitian conjugation is not a unitary operator (similar to other matrix-manipulations in \cite{ZQKW_2024,ZBQKW_arXive2024}), whose realization becomes possible due to its application to the matrix encoded into the quantum state in a specific way.
(ii) Another issue addressed in this paper is weakening the normalization constraint for the elements of the matrices { encoded into  quantum states}. We recall that such constraint is  imposed by the normalization condition for a pure quantum state. For that  purpose,  we introduce an additional probability amplitude into the superposition state encoding considered matrix. This can be done involving one more  auxiliary one-qubit subsystem.
{
Both of the above modifications are applicable to the algorithms involving { encoding of  particular quantum states}. For instance, the already mentioned HHL-algorithm \cite{HHL} uses  encoding the known right-hand side of the linear system,   the matrix-manipulation algorithms in  \cite{ZZRF} encode the vectors subjected to the matrix operations,  the algorithms in \cite{ZQKW_2024,ZBQKW_arXive2024} encode  the matrices for calculating addition, multiplication, inversion,  quantum machine learning \cite{BWPRWL,A,RML} requires encoding the input data,  encoding the  large data sets is utilized in the least-square linear-regression algorithms  \cite{WBL,SSP,Wang2}  and so on.}

Of course, the QCM and both above extensions require modifications of the mentioned algorithms, in particular, the algorithms  proposed in \cite{ZQKW_2024,ZBQKW_arXive2024}.
Below we present the detailed analysis of the appropriate modifications introduced into  the  multiplication algorithm (see appendix in \cite{ZBQKW_arXive2024}) and demonstrate manipulations with the input matrices   involving  Hermitian conjugation. { Of course,  similar modifications  can be implemented in other matrix-manipulation algorithms such as inner product, addition, determinant calculation, matrix inversion, linear system solving.} All quantum circuits associated with QCM  and  two above extensions, as well as  the circuit for modified multiplication algorithm,  are  included.

The paper is organized as follows. The QCM  is proposed in Sec.\ref{Section:CMA} with all necessary details.  In Sec.\ref{Section:applications}, we consider applications of QCM  to algorithms of matrix manipulations. Then,  in Sec.\ref{Section:extension}, we introduced two pointed above  extensions to the matrix-manipulation algorithms. In Sec.\ref{Section:mult}, we implement the QCM  together with two above  extensions into the matrix multiplication algorithm.  { In the same section,} we demonstrate possibility of manipulating the input matrices using subroutine of Hermitian conjugation and show how to measure the output normalization constant   that is lost because of the QCM  
of the ancilla state at the last step of the algorithm. Conclusions are given in Sec.\ref{Section:conclusions}. {    The particular example of matrix multiplication is considered in the Appendix, Sec.\ref{Section:Appendix}.}

\section{QCM in quantum algorithms}
\label{Section:CMA}
\subsection{Motivation}
\label{Section:M}
{  As mentioned in the Introduction}, the necessity for introducing the QCM instead  of the usual one is motivated by small success probability in the  ancilla-state measurement that appears, for instance, in quantum algorithms for  matrix manipulations  \cite{ZQKW_2024,ZBQKW_arXive2024}. Small success probability requires multiple runs of the algorithm to get the desired  result of measurement. The QCM completely removes  this obstacle to effective application of  the above algorithms.

As a simple example motivating this study, let us consider the superposition quantum state
 of $n^{(S)}$ qubit subsystem $S$ and $n^{(R)}$ qubit subsystem $R$.
The state of the whole system $S\otimes R$ reads
\begin{eqnarray}\label{state1}
|\Phi_0\rangle =\sum_{i=0}^{N^{(S)}-1} \sum_{j=0}^{N^{(R)}-1} a_{ij}|i\rangle_S |j\rangle_R ,\;\;\sum_{i=0}^{N^{(S)}-1} \sum_{j=0}^{N^{(R)}-1} |a_{ij}|^2=1,\;\;N^{(S)}=2^{n^{(S)}}, \;\; N^{(R)}=2^{n^{(R)}}.
\end{eqnarray}
Our purpose is to extract the (normalized) sum of $a_{ij}$ over the second subscript, i.e. we need to  construct the state
\begin {eqnarray}\label{state2}
\sim \sum_{i=0}^{N^{(S)}-1} \left(\sum_{j=0}^{N^{(R)}-1}a_{ij}\right) |i\rangle_S.
\end{eqnarray}
This task, { for instance,} is typical for the algorithms proposed in    \cite{ZQKW_2024,ZBQKW_arXive2024} and therefore deserves the special consideration.

To transfer state in Eq.(\ref{state1}) to state in Eq.(\ref{state2}), we, first,
apply the Hadamard transformation $H$ to each qubit of $R$ collecting all $H$   in the operator $H_R=H^{\otimes N^{(R)}}$:
\begin{eqnarray}\label{Phi1}
|\Phi_1\rangle=H_R |\Phi_0\rangle =\frac{1}{2^{n^{(R)}/2}} \sum_{i=0}^{N^{(S)}-1} \sum_{j=0}^{N^{(R)}-1} a_{ij}|i\rangle_S |0\rangle_R
  + |g\rangle_{SR}.
\end{eqnarray}
Hereafter the subscript at the operator means the subsystem to which this operator is applied.
All needed terms are collected in the first part of Eq.(\ref{Phi1}), while  all the rest terms are collected in the garbage  unnormalized state $|g\rangle_{SR}$, orthogonal to the basis states in the first term of $|\Phi_1\rangle$.
Now, to label the garbage, we introduce the one-qubit ancilla $B_1$ in the ground state and apply the controlled operator
\begin{eqnarray}\label{WRA}
W^{(1)}_{RB_1}=|0\rangle_R \, _R\langle 0| \otimes \sigma^{(x)}_{B_1} + (I_R -  |0\rangle_R \, _R\langle 0| ) \otimes I_{B_1}
\end{eqnarray}
to the state
$|\Phi_1\rangle |0\rangle_{B_1} $:
\begin{eqnarray}\label{chi2}
|\Phi_2\rangle=W^{(1)}_{RB_1}|\Phi_1\rangle |0\rangle_{B_1} =\frac{1}{2^{n^{(R)}/2}} \sum_{i=0}^{N^{(S)}-1} \sum_{j=0}^{N^{(R)}-1} a_{ij}|i\rangle_S |0\rangle_R |1\rangle_{B_1} + |g\rangle_{SR}|0\rangle_{B_1}.
\end{eqnarray}
Finally,  we can remove the garbage via ancilla measurement with the desired output { $|1\rangle_{B_1}$} thus obtaining the state
\begin{eqnarray}\label{Phi3}
&&
|\Phi_3\rangle=|\Psi_{out}\rangle_S |0\rangle_R { |0\rangle_{B_1}} , \;\; \\\label{G0}
&& |\Psi_{out}\rangle_S=G^{-1}\sum_{i=0}^{N^{(S)}-1} \sum_{j=0}^{N^{(R)}-1} a_{ij}|i\rangle_S, \;\;
G=\sqrt{\sum_{i=0}^{N^{(S)}-1}\left| \sum_{j=0}^{N^{(R)}-1} a_{ij}\right|^2}.
\end{eqnarray}
However, the probability of  obtaining the desired state { $|1\rangle_{B_1}$ }of the ancilla $B_1$ as the result of ancilla measurement (success probability)  is $p=\frac{G^2}{2^{n^{(R)}}}\sim \frac{1}{2^{n^{(R)}}}$
and decreases exponentially with $n^{(R)}$.  {     This means that the number of runs needed to access the desired ancilla state can be estimated as $O\left( \frac{2^{n^{(R)}}}{G^2}\right)$. { We also notice that, if the only required result is the state   $|\Phi_3\rangle$, then the single
successful measurement of the ancilla state  is enough to end up with this state of the subsystem $S\otimes R$.  However, if  we also need to calculate the normalization $G$, then the single successful measurement is not enough because $G$ can be calculated only probabilistically and the set of runs is requires to determine the success probability $p$, so that  $G=2^{n^{(R)}/2} \sqrt{p}$. }

Thus, in general, the proposed algorithm of selecting the needed state $|\Phi_3\rangle$ (post-selection)  becomes  ineffective requiring the large number of running to succeed in measuring the desired state  $|1\rangle_{B_1}$.  This statement motivates the search for  methods allowing to overcome this obstacle removing the problem of small success probability.  Such method is proposed in the next section,  Sec.\ref{Section:contrmeas}.

\subsection{QCM  and its features}
\label{Section:contrmeas}
{ In this section we extend the concept of measurement as a tool allowing to get   information from the quantum state of some system. First we  recall the  concept of measurement   restricting  ourselves to the one-qubit measurements \cite{NCh}.  The usage of measurement in our manuscript is in removing the garbage states, i.e., the states which appear as a byproduct of a quantum algorithm, see, for instance, \cite{HHL}. Then we introduce the  one-qubit QCM as a tool allowing to  avoid the small success probability revealed in Sec.\ref{Section:M}, thus removing the necessity of multiple running of the algorithm with the purpose to register the desired ancilla state. As a consequence, QCM makes the garbage removing  much more effective.   Thus,  QCM becomes the preferable method for garbage removing and its application is demonstrated  in Secs.\ref{Section:applications} and \ref{Section:mult}.

\subsubsection{One-qubit measurement}
\label{Section:oneq}
Hereafter, by the measurement (projective measurement) of the  one-qubit system $B_1$   we call the operator $M_{B_1}$ that, acting on the superposition of basis states
\begin{eqnarray}\label{oneq}
|\Phi_0\rangle_{B_1} =\alpha |1\rangle_{B_1} + \beta |0\rangle_{B_1},\;\; |\alpha|^2 +  |\beta|^2=1,
\end{eqnarray}
{ transfers it to one  of the basis states, either  $|1\rangle_{B_1}$ or $|0\rangle_{B_2}$, with the probability, respectively, $|\alpha|^2$ and $|\beta|^2$, i.e.,
 \begin{eqnarray}
 M_{B_1} |\Phi_0\rangle_{B_1} =\left\{
 \begin{array}{ll}\displaystyle
  |1\rangle_{B_1}, & {\mbox{probability}} \;\; |\alpha|^2\cr\displaystyle
|0\rangle_{B_1}, & {\mbox{probability}} \;\; |\beta|^2
  \end{array}
 \right. .
 \end{eqnarray}
 Therefore, applying $M_{B_1}$ to the  state $\alpha |\varphi_1 \rangle |1\rangle_{B_1} + \beta |\varphi_2 \rangle |0\rangle_{B_1}$, where $|\varphi_j\rangle $, $j=1,2$, are some normalized states, yields
  \begin{eqnarray}
 M_{B_1}\Big(\alpha |\varphi_1 \rangle |1\rangle_{B_1} + \beta |\varphi_2 \rangle |0\rangle_{B_1}\Big)=\left\{
 \begin{array}{ll}\displaystyle
 \frac{\alpha}{|\alpha|}  |\varphi_1\rangle|1\rangle_{B_1}, & {\mbox{probability}} \;\; |\alpha|^2\cr\displaystyle
  \frac{\beta}{|\beta|}  |\varphi_2\rangle |0\rangle_{B_1}, & {\mbox{probability}} \;\; |\beta|^2
  \end{array}
 \right. .
 \end{eqnarray}}
 Thus, the qubit after  measurement can not remain in  a superposition state. The property of the measurement operator to select a particular basis states serves to remove the extra terms (called garbage) from the above quantum state, given in Eq.(\ref{chi2}),  reducing it  to Eq.(\ref{Phi3}), see also \cite{HHL,ZQKW_2024,ZBQKW_arXive2024}. We emphasize that the measurement becomes deterministic if either $\alpha=0$ or $\beta=0$ in the superposition  quantum state (\ref{oneq}).

\subsubsection{One-qubit QCM}
\label{Section:CM}
Now, along with the system $B_1$, we introduce the one-qubit ancilla  $B_2$ in the ground state $|0\rangle_{B_2}$ and apply the operator  C-NOT
\begin{eqnarray}\label{cnot}
W^{(2)}_{B_1B_2}=|1\rangle_{B_1} \, _{B_1}\langle 1| \otimes \sigma^{(x)}_{B_2} + |0\rangle_{B_1} \, _{B_1}\langle 0| \otimes I_{B_2}
\end{eqnarray}
to  the state $|\Phi_0\rangle_{B_1} |0\rangle_{B_2}$:
\begin{eqnarray}\label{W2Phi0}
|\Phi_1\rangle_{B_1B_2} =W^{(2)}_{B_1B_2}|\Phi_0\rangle_{B_1} |0\rangle_{B_2} = \alpha |1\rangle_{B_1} |1\rangle_{B_2}+ \beta |0\rangle_{B_1} |0\rangle_{B_2},
\end{eqnarray}
thus doubling the state of $B_1$.
Next, we  introduce the following operator, called   the quantumly controlled measurement, using $B_1$ as the controlling
qubit and $B_2$ as the controlled qubit:
\begin{eqnarray}\label{CM}
W^{(3)}_{B_1B_2} = |1\rangle_{B_1} \, _{B_1}\langle 1| \otimes M_{B_2} + |0\rangle_{B_1} \, _{B_1}\langle 0|\otimes I_{B_2},
\end{eqnarray}
where $M_{B_2}$ is the measurement operator, discussed in Sec.\ref{Section:oneq}, applied to the ancilla $B_2$. Notice that the operator $W^{(3)}_{B_1B_2}$ is not a unitary one due to the measurement $M_{B_2}$. Moreover, the structure of this operator is unusual  because the measurement operator is never used as a controlled operator in the standard controlled-operator technique.
Applying this operator  to $|\Phi_1\rangle_{B_1B_2}$ we obtain  the state
\begin{eqnarray}\label{Mf}
W^{(3)}_{B_1B_2}|\Phi_1\rangle_{B_1B_2} =  \alpha M_{B_2}  |1\rangle_{B_1} |1\rangle_{B_2}+\beta |0\rangle_{B_1}|0\rangle_{B_2}.
\end{eqnarray}
 We emphasize that Eq.(\ref{Mf}) has an unusual structure because the measurement is applied to the fist term of the superposition state only, while usually the measurement can be applied to the whole superposition state. This selective application of measurement is a consequence of the quantum control in the operator $M_{B_2}$.  Eq.(\ref{Mf}) demonstrates that the measurement of $B_2$ is switched on by the  state  $|1\rangle_{B_1}$ of $B_1$ and the operator $M_{B_2}$ is applied to the desired  state $|1\rangle_{B_2}$. Therefore, if $\alpha\neq 0$, the measurement deterministically  yields only $|1\rangle_{B_2}$ without the choice between  $|0\rangle_{B_2}$  and $|1\rangle_{B_2}$. In other words, we deal with the perfect success in getting the desired state $|1\rangle_{B_2}$, there is no post-selection.  This means that the second term must disappear from (\ref{Mf}) because it includes state $|0\rangle_{B_2}$ rather than $|1\rangle_{B_2}$.  Thus
\begin{eqnarray}\label{Mf2}
W^{(3)}_{B_1B_2}|\Phi_1\rangle_{B_1B_2} =  \left\{ \begin{array}{ll} \displaystyle
\frac{\alpha}{|\alpha|} |1\rangle_{B_1} |1\rangle_{B_2},& \alpha\neq 0\cr\displaystyle
\frac{\beta}{|\beta|}  |0\rangle_{B_1} |0\rangle_{B_2},&  \alpha=0
\end{array}\right. .
\end{eqnarray}
Similarly, the QCM can be applied to the state
\begin{eqnarray}\label{tMf}
|\tilde \Phi_1\rangle =  \alpha |E_1\rangle_S  |1\rangle_{B_1} |1\rangle_{B_2}+\beta  |E_2\rangle_S  |0\rangle_{B_1}|0\rangle_{B_2},\;\; |\alpha|^2 + |\beta|^2=1,
\end{eqnarray}
where $|E_i\rangle_S$, $i=1,2$, are some normalized states of the system $S$, yielding the following result:
\begin{eqnarray}\label{tMf2}
W^{(3)}_{B_1B_2}|\tilde \Phi_1\rangle=  \left\{ \begin{array}{ll} \displaystyle
\frac{\alpha}{|\alpha|}|E_1\rangle_S  |1\rangle_{B_1} |1\rangle_{B_2},& \alpha\neq 0\cr\displaystyle
\frac{\beta}{|\beta|} |E_2\rangle_S  |0\rangle_{B_1} |0\rangle_{B_2},&  \alpha=0
\end{array}\right. .
\end{eqnarray}
 We emphasize that the states of $B_2$ in the first and second terms of the state $|\Phi_1\rangle_{B_1B_2}$ in Eq.(\ref{W2Phi0}) are orthogonal to each other, so that the second term of Eq.(\ref{W2Phi0}) does not contribute to the output in Eq.(\ref{Mf2})  after QCM  if $\alpha\neq 0$.

 {
 However, the different situation is also possible.
To reveal it,
we apply QCM $W^{(3)}_{B_1B_2}$ to the state $|\Phi_1\rangle_{B_1B_2} = |\Phi_0\rangle_{B_1} |0\rangle_{B_2}$. Instead of (\ref{Mf}), we have
\begin{eqnarray}\label{Mff}
W^{(3)}_{B_1B_2}|\Phi_1\rangle_{B_1B_2} =  \alpha M_{B_2}  |1\rangle_{B_1} |0\rangle_{B_2}+\beta |0\rangle_{B_1}|0\rangle_{B_2}.
\end{eqnarray}
Again, the measurement is initiated by the first term of Eq.(\ref{Mff})  and fixes  the state $|0\rangle_{B_2}$ of the qubit $B_2$. But  the same state  $|0\rangle_{B_2}$ appears in the second term of Eq.(\ref{Mff}).  Consequently, this term also contributes to the resulting state . Therefore, the result of QCM is
\begin{eqnarray}\label{Mff2}
W^{(3)}_{B_1B_2}|\Phi_1\rangle_{B_1B_2} = \Big(\alpha|1\rangle_{B_1}+\beta |0\rangle_{B_1}\Big)|0\rangle_{B_2}.
\end{eqnarray}
This is the case when the quantumly controlled measurement is equivalent to the usual one.  }
 Again, we emphasize the unusual  structure of Eq. (\ref{Mff}) where the measurement operator is applied to the first term only. This makes the measurement deterministic putting  $B_2$ to the state $|0\rangle_{B_2}$ and therefore, since the same state $|0\rangle_{B_2}$ is present in the second term,  leading to the final state in Eq. (\ref{Mff2}).  

Formulae (\ref{Mf2}) and (\ref{Mff2}) can be generalized as follows. We assume that the controlled state of  $B_2$  is a superposition state, i.e., we replace the state $|\Phi_1\rangle_{B_1B_2}$ given in Eq.(\ref{W2Phi0}) with the following one:
\begin{eqnarray}\label{NPhi}
|\Phi_1\rangle_{B_1B_2} =  \alpha |1\rangle_{B_1} (\alpha_1 |0\rangle_{B_2} + \beta_1|1\rangle_{B_2}) +  \beta |0\rangle_{B_1} (\alpha_2 |0\rangle_{B_2} + \beta_2|1\rangle_{B_2}) ,\;\; |\alpha_i|^2+|\beta_i|^2=1,\;\;i=1,2.
\end{eqnarray}
We also assume  that the superposition states of $B_2$ used in the first and second terms of Eq.(\ref{NPhi}) are not orthogonal  to each other in general.
Instead of Eq.(\ref{CM}), we have
\begin{eqnarray}\label{CM0}
W^{(3)}_{B_1B_2}|\Phi_1\rangle_{B_1B_2} =  \alpha |1\rangle_{B_1} M_{B_2} (\alpha_1 |0\rangle_{B_2} + \beta_1|1\rangle_{B_2}) +  \beta |0\rangle_{B_1} (\alpha_2 |0\rangle_{B_2} + \beta_2|1\rangle_{B_2}) .
\end{eqnarray}
Then, the measurement $M_{B_2}$ initiated by the first term in (\ref{CM0})   yields the state $|0\rangle_{B_2}$ with the probability $p_1=|\alpha_1|^2$ and the state $|1\rangle_{B_2}$ with the probability $p_2=|\beta_1|^2$ { (because the measurements is applied to the state of $B_2$ associated with the first term of Eq.(\ref{CM0})),} both these probabilities do not depend on $\beta$. Now, to correctly write the resulting state after applying $W^{(3)}_{B_1B_2}$, we emphasize that the state  $|1\rangle_{B_1}$ of $B_1$ just switches on the measurement $M_{B_2}$ thus destroying the superposition state of $B_2$ putting it to the basis  state, either $|0\rangle_{B_2}$ or $|1\rangle_{B_2}$. However, the fact that the state becomes fixed holds for both terms of Eq.(\ref{CM0}) (similar to Eq.(\ref{Mff2}))
 because the qubit  $B_2$ can not remain in a superposition state after measurement.  Thus, since both terms in (\ref{CM0}) contain two basis states $|0\rangle_{B_2} $ and $|1\rangle_{B_2}$,  both of these terms yield contributions to the resulting state of quantum system after QCM.
Therefore, we have
{
\begin{eqnarray}\label{res2}
&&
W^{(3)}_{B_1B_2} |\Phi_1\rangle_{B_1B_2} = \\\nonumber
&&\left\{
\begin{array}{ll}
G_1^{-1}\Big(\alpha \alpha_1| 1\rangle_{B_1}  + \beta\alpha_2 |0\rangle_{B_1}\Big)    |0\rangle_{B_2}, & \alpha \neq 0,\;\; \displaystyle  M_{B_2}   (\alpha_1 |0\rangle_{B_2} + \beta_1|1\rangle_{B_2})  =
\frac{\alpha_1}{|\alpha_1|} |0\rangle_{B_2},\;\;  p_1=|\alpha_1|^2\cr
G_2^{-1}\Big(\alpha \beta_1| 1\rangle_{B_1}  + \beta \beta_2 |0\rangle_{B_1}\Big) { |1\rangle_{B_2}}, & \alpha \neq 0,\;\; \displaystyle   M_{B_2}  (\alpha_1 |0\rangle_{B_2} + \beta_1|1\rangle_{B_2}) = {  \frac{\beta_1}{|\beta_1|}  |1\rangle_{B_2}},\;\; p_2=|\beta_1|^2\cr
 e^{i\varphi_\beta} | 0\rangle_{B_1}(\alpha_2 |0\rangle_{B_2} + \beta_2|1\rangle_{B_2}) ,& \alpha=0, \;\;\beta =e^{i\varphi_\beta},\;\;0\le \varphi_\beta \le 2\pi
\end{array}
\right.,\\\label{GG}
&&
G_1=\sqrt{|\alpha\alpha_1|^2 + |\beta\alpha_2|^2}, \;\;G_2=\sqrt{|\alpha\beta_1|^2 + |\beta\beta_2|^2}.
\end{eqnarray}
Thus, we have a  superposition state in all three cases.
   If $\alpha=0$, then the operator $W^{(3)}_{B_1B_2}$ does not apply the measurement $M_{B_2}$ as shown in the third line in the right-hand side of Eq.(\ref{res2}).
   We note that applying the usual measurement to the state $|\Psi_1\rangle_{B_1B_2}$ yields
\begin{eqnarray}\label{res2M}
M_{B_2} |\Phi_1\rangle_{B_1B_2} = \left\{
\begin{array}{ll}
G_1^{-1}\Big(\alpha \alpha_1| 1\rangle_{B_1}  + \beta\alpha_2 |0\rangle_{B_1}\Big)    |0\rangle_{B_2}, &p_1=G_1^2\cr
G_2^{-1}\Big(\alpha \beta_1| 1\rangle_{B_1}  + \beta \beta_2 |0\rangle_{B_1}\Big) { |1\rangle_{B_2}}, & p_2=G_2^2
\end{array}
\right.,
\end{eqnarray}
i.e.,
 the same states as in  the first and second lines in the right-hand side of Eq.(\ref{res2}), but the probabilities of obtaining these  states are different  and equal to the probabilities of measuring the states $|0\rangle_{B_2}$ and $|1\rangle_{B_2}$ respectively.
   }

{
\subsection{Properties of QCM}
\label{Section:properties}
We collect some most important properties of QCM.
\begin{enumerate}
\item
To access the desired state of the ancilla via the usual measurement the multiple run of the algorithm is required, whereas the single run is enough to do the same  via QCM.
As a  consequence, QCM does not supply any classical information, unlike  the ordinary measurement. In fact, we do not obtain the normalization $G$ in Eq.(\ref{G0}) in terms of  the  probability $p$, because the result of measurement of the ancilla $B_2$  is predictable and equals $|1\rangle_{B_2}$.  Thus, the quantum state obtained after QCM can not be considered as a result of post-selection.  
\item
QCM $W^{(3)}_{B_1B_2}$, defined in Eq.(\ref{CM}), is not a unitary operator due to the presence of measurement $M_{B_2}$.
\item
{
The definition  of QCM, given in  Eq.(\ref{CM}), can be generalized as follows:
\begin{eqnarray}\label{CM2}
W^{(3)}_{B_1B_2} = |1\rangle_{B_1} \, _{B_1}\langle 1| \otimes M_{B_2} + |0\rangle_{B_1} \, _{B_1}\langle 0|\otimes \tilde M_{B_2},
\end{eqnarray}
where $\tilde M_{B_2}$ is another  measurement  applied to $B_2$.  If $\tilde M_{B_2} \equiv M_{B_2}$, then
applying this operator to the state $|\Phi_1\rangle_{B_1B_2}$ given in (\ref{W2Phi0}), we obtain
\begin{eqnarray}
W^{(3)}_{B_1B_2}  |\Phi_1\rangle_{B_1B_2} =\left\{
\begin{array}{ll}\displaystyle
\frac{\alpha}{|\alpha|} |1\rangle_{B_1}  |1\rangle_{B_2}, &p_1= |\alpha|^2\cr
\displaystyle\frac{\beta}{|\beta|} |0\rangle_{B_1}  |0\rangle_{B_2}, &p_2= |\beta|^2
\end{array}
\right..
\end{eqnarray}}
In this case, the result of applying QCM coincides with the result of standard measurement.
\item
QCM maps the pure state into another pure state, therefore it is a trace-preserving  positive map.
\item
Although QCM is  seemingly equivalent to  the amplification of the probability   of obtaining the desired state, it is  not a variant/generalization of the amplitude amplification algorithm based on  the Grover's search algorithm \cite{NCh,BHMP,KB,Liu_2024,KS_2025,CZS}. The principal  difference is including the measurement (switched on by the quantum tool) in the operator of QCM. Therefore we deal with the operator combining both quantum and classical features.  Unlike the amplitude amplification via the modifications of the Grover's algorithm, which require multiple applying of  certain unitary operator, QCM  assumes the single run. {More detailed comparison of QCM  and amplitude amplification is give n in Sec.\ref{Section:CMandQAA}.}
\item
Application of QCM  to the density matrix is also  possible. Details are left for further study.
\end{enumerate}

We also note that including QCM  requires only one-qubit additional ancilla $B_2$ and therefore can not be considered as a space-consuming operator.

\subsection{Quantumly controlled projector}

The quantumly controlled projector (QCP) can be introduced
by the formula
\begin{eqnarray}\label{CP}
\tilde W^{(3)}_{B_1B_2} = |1\rangle_{B_1} \, _{B_1}\langle 1| \otimes P_{B_2} + |0\rangle_{B_1} \, _{B_1}\langle 0|\otimes I_{B_2},
\end{eqnarray}
where $P_{B_2}$ is some projector. For instance, in the computational basis this projector can be either
$|1\rangle_{B_2}\, _{B_2}\langle 1|$ or
$|0\rangle_{B_2}\, _{B_2}\langle 0|$. This operator also resolves the problem of small success probability and might be simpler for realization than QCM.
It is obvious that the action of $\tilde W^{(3)}_{B_1B_2}$ on $|\Phi_1\rangle_{B_1B_2}$ in Eq.(\ref{W2Phi0}) coincides with the action of
$W^{(3)}_{B_1B_2}$ on the same state, see  Eq. (\ref{Mf2}).
Although  the properties of QCP differ from the properties of QCM presented in Sec.\ref{Section:properties}, they  can be written down in a quite similar way.
}

{\subsection{QCM and quantum amplitude-amplification}
\label{Section:CMandQAA}
Although QCM  seems to solve the same problem as quantum  amplification  algorithm, there is a principal difference in realization. Quantum amplification algorithms uses only unitary transformations, while representation of QCM  in terms of unitaries is doubtful because of presence of measurement operator.

In principle, we can represent QCM  in terms of oracle operators as follows.
Let us start with the state
\begin{eqnarray}
|\Psi\rangle = \alpha |E_1\rangle + \beta |E_2\rangle, \;\;\langle E_i | E)j \rangle = \delta_{ij}.
\end{eqnarray}
We would like to end up with the state   $|E_1\rangle$ up to some phase factor:
\begin{eqnarray}\label{CMpsiout}
|\Psi_{out}\rangle =e^{i \varphi}  |E_1\rangle, \;\; 0\le \varphi < 2 \pi,
\end{eqnarray}
i.e., $|E_2\rangle$ is garbage.
We attach the one-qubit uncilla $B_1$ in the ground state to $|\Psi\rangle$ and apply the Grover's oracle operator, assuming that the oracle knows the required state $|E_1\rangle$, so that the oracle action yields
\begin{eqnarray}\label{Phi110}
|\Phi_1\rangle= O_G |\Psi\rangle |0\rangle_{B_1} =  \alpha |E_1\rangle  |f(E_1)\rangle_{B_1}+ \beta |E_2\rangle |f(E_2)\rangle_{B_1}=
  \alpha |E_1\rangle  |1\rangle_{B_1}+ \beta |E_2\rangle |0\rangle_{B_1}
,
\end{eqnarray}
where $f(E_1)=1$ and $f(E_2)=0$, similar to the usual Grover's algorithm.
Namely this step is realized in Eq.(\ref{chi2}).

Next, we attach another one-qubit ancilla $B_2$ in the ground state and apply C-NOT $W^{(2)}_{B_1B_2}$ given in
(\ref{cnot}):
\begin{eqnarray}\label{Phi20}
|\Phi_2\rangle=W^{(2)}_{B_1B_2}|\Phi_1\rangle |0\rangle_{B_2} =   \alpha |E_1\rangle  |1\rangle_{B_1}  |1\rangle_{B_2}+ \beta |E_2\rangle |0\rangle_{B_1}  |0\rangle_{B_2}.
\end{eqnarray}
Now, instead of using the unitary transformation to rotate the vector-state over some angle $\theta$, like in the Grover's algorithm, we involve another QCM-oracle (i.e., the oracle which is able to identify the state
$|1\rangle_{B_1}$ in the superposition state $|\Phi_2\rangle$ and  perform QCM, thus fixing  $B_2$ in the state $|1\rangle_{B_2}$):
\begin{eqnarray}
|\Phi_3\rangle=O_{QCM} |\Phi_2\rangle  = |\Psi_{out}\rangle |1\rangle_{B_1} |1\rangle_{B_2}
\end{eqnarray}
with $ |\Psi_{out}\rangle$ given in  Eq.(\ref{CMpsiout})  with $\varphi = \arg \alpha$.
Thus, in terms of oracles, we apply  three operators to the original state $|\Psi\rangle |0\rangle_{B_1}|0\rangle_{B_2}$:
\begin{eqnarray}\label{Phi31}
|\Phi_3\rangle =O_{QCM} W^{(2)}_{B_1B_2} O_G|\Psi\rangle |0\rangle_{B_1}|0\rangle_{B_2}.
\end{eqnarray}
All three operations compose the oracle-interpretation of the controlled measurement and  are shown in Fig.\ref{Fig:QAACM}.

\begin{figure}[ht]
    \includegraphics[width=0.4\textwidth]{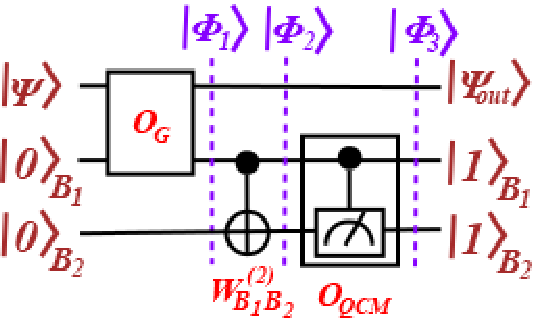}
    \caption{Oracle-representation of the QCM subroutine. The intermediate states $|\Phi_1\rangle$, $|\Phi_2\rangle$ and $|\Phi_3\rangle$ are determined, respectively, in Eqs. (\ref{Phi110}), (\ref{Phi20}) and (\ref{Phi31}). }
\label{Fig:QAACM}
\end{figure}
}

\subsection{ On realizability of QCM}
	 { There is no principal objections to realizability of  QCM.  Due to  the presence of the  state $|1\rangle_{B_1}$ (which switches on the measurement $M_{B_2}$ of the state $|1\rangle_{B_2}$)  in the  superposition  state  $|\Phi_1\rangle_{B_1B_2}$ (see Eqs.(\ref{W2Phi0}), (\ref{NPhi})), we do not  need to repeat running the algorithm even if  the probability of access to the desired state  $|1\rangle_{B_2}$ via the usual measurement  is small.} This is the main difference between  QCM and the usual measurement where the value of access probability to the desired ancilla-state   has the   crucial meaning.   In other words, we switch on the  measurement {  (that can be called a classical operation)} by the quantum  tool  (via the state  $|1\rangle_{B_1}$)  relying on reality of quantum superposition states. Therefore, realizability of  QCM would additionally
	confirm the existence of such states.
	
The probabilistic nature of the standard measurement comes from the interplay between the classical and quantum operations governed by the classical principles of extracting information from the quantum system.
Therefore, the usual measurement can be referred to the classical-classical controled operator, which  switch on the measurement ``by hand'' whenever the observer (the classical controlling operator)  decides to get the classical information out of the quantum state. Although  the classical controlling observer is not indicated in standard quantum circuits.
 In  QCM, there is also classical-quantum interplay, but it is governed by the quantum principles and therefore is a new type  of control. In particular, the quantum control of measurement may completely remove its probabilistic feature (see eq.(\ref{Mf2}))  which is the  natural feature of the usual measurement.
	Thus,  QCM can be considered as  applying the classical operator (measurement) to some subsystem (ancilla $B_2$) controlled  by the quantum state of another subsystem (ancilla $B_1$).
	 Notice that the opposite control  (i.e., applying  the quantum operator controlled by the classical state of some subsystem, { i.e., by the state after measurement}) is an acknowledged controlled  operation used, for instance, in the teleportation algorithm    \cite{BBCJPW,NCh}, this is the classical-quantum control.  The quantum-quantum control is also well known, C-NOT is the simplest representative of such operation. In this case both controlling and controlled subsystems and  controlled operator are quantum ones.  Therefore, realization of  QCM
	 would be an example of the fourth type of control in the hybrid quantum-classical system and can be referred to as  the  quantum-classical control.
	At the moment, we can not suggest a particular realization of  QCM in terms of the well-known quantum and classical operators. However, its  advantage over the usual measurement   demonstrated above stimulates the search for  methods realizing  such operator.

\section{Applications of QCM}
\label{Section:applications}
\subsection{{  QCM in algorithm developed in Sec.\ref{Section:M}}}
 { Let us demonstrate applications of   QCM in the algorithm developed in Sec.\ref{Section:M}}.} We include the one more one-qubit ancilla $B_2$ in the ground state and apply C-NOT $W^{(2)}_{B_1B_2}$ , given in Eq.(\ref{cnot}),
to $|\Phi_2\rangle |0\rangle_{B_2}$ { (with $|\Phi_2\rangle$ defined in Eq.(\ref{chi2}))} thus  doubling  the state of the ancilla $B_1$:
\begin{eqnarray}\label{Phi33}
|\Phi_3\rangle=W^{(2)}_{B_1B_2}|\Phi_2\rangle |0\rangle_{B_2} =\frac{1}{2^{n^{(R)}/2}} \sum_{i=0}^{N^{(S)}-1} \sum_{j=0}^{N^{(R)}-1} a_{ij}|i\rangle_S |0\rangle_R
|1\rangle_{B_1}|1\rangle_{B_2} + |g\rangle_{SR}|0\rangle_{B_1}|0\rangle_{B_2}.
\end{eqnarray}
{ Next, we apply  $W^{(3)}_{B_1B_2}$, defined in Eq.(\ref{CM}), to $|\Phi_3\rangle$ and obtain, according to Eq.(\ref{tMf2}),
\begin{eqnarray}\label{Phi43}
|\Phi_4\rangle=W^{(3)}_{B_1B_2}|\Phi_3\rangle =\frac{1}{2^{n^{(R)}/2}} \sum_{i=0}^{N^{(S)}-1} \sum_{j=0}^{N^{(R)}-1} a_{ij}|i\rangle_S |0\rangle_R
\Big(M_{B_2}|1\rangle_{B_1}|1\rangle_{B_2}\Big) + |g\rangle_{SR}|0\rangle_{B_1}|0\rangle_{B_2}.
\end{eqnarray}
 Using Eq.(\ref{tMf2}) with $\alpha \neq 0$,  we  stay with the first  term involving $|1\rangle_{B_2}$ in (\ref{Phi33}) thus removing the garbage} and resulting in
\begin{eqnarray}\label{chi4}
|\Phi_4\rangle=
|\Psi_{out}\rangle_S |0\rangle_R |1\rangle_{B_1} { |1\rangle_{B_2}} ,
\end{eqnarray}
{ where $|\Psi_{out}\rangle_S$ and the  normalization constant $G$ are  defined in  Eq.(\ref{G0}).}
This { modification of} the  algorithm  allows us to avoid the problem of small success probability for the ancilla-state measurement. It is remarkable that we do not even need to read the result of measurement because it is known in advance and, moreover, we do not use this result of measurement. This operation is involved just to remove the garbage from the superposition quantum state.
The depth of the described algorithm is determined by the operator $W^{(1)}_{RB_1}$ and equals $O(n^{(R)})$.
The  circuit realizing this algorithm is given in Fig.\ref{Fig:CM2}.

\begin{figure}[ht]
    \includegraphics[width=0.4\textwidth]{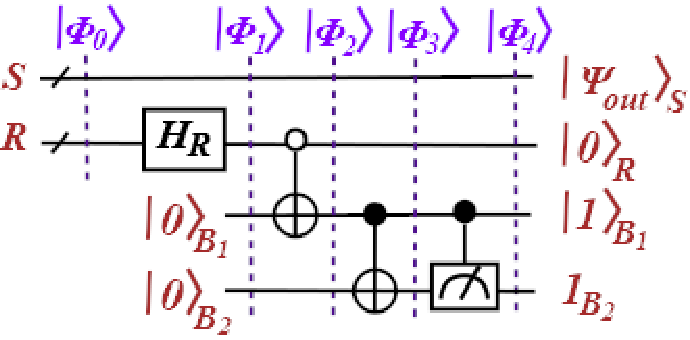}
    \caption{Application of QCM {to the algorithm present in Sec.\ref{Section:M}}. The intermediate states $|\Phi_j\rangle$, $j=0,\dots,4$, are determined, respectively in Eqs. (\ref{state1}), (\ref{Phi1}), (\ref{chi2}), (\ref{Phi33}) and (\ref{chi4}). } 
\label{Fig:CM2}
\end{figure}

\subsection{ QCM in algorithms present in Refs.\cite{ZQKW_2024,ZBQKW_arXive2024}}
All the states obtained in Refs.\cite{ZQKW_2024,ZBQKW_arXive2024} before applying the final ancilla-state measurement have the form
\begin{eqnarray}\label{EF}
|\Phi\rangle = \alpha |E\rangle |1\rangle_{B_1} + \beta |F\rangle |0\rangle_{B_1}, \;\; |\alpha|^2+|\beta|^2 =1,\;\;|\alpha|\ll |\beta|,\;\;\langle F|E\rangle=0, \;\; \langle E|E\rangle=\langle F|F\rangle=1.
\end{eqnarray}
where the state $|E\rangle $ contains  the useful information, while the state  $|F\rangle$ gathers all the garbage. In Eq.(\ref{EF}), the useful information and garbage are labeled by the states, respectively, $|1\rangle_{B_1}$ and $|0\rangle_{B_1}$ of the one-qubit ancilla $B_1$.
To arrange  QCM we, first of all,  introduce the additional one-qubit ancilla $B_2$ in the ground state  and apply the operator $W^{(2)}_{B_1B_2}$, defined in Eq.(\ref{cnot}),
to the state $|\Phi\rangle|0\rangle_{B_2}$:
\begin{eqnarray}\label{SPhi1}
|\Phi_1\rangle = W^{(2)}_{B_1B_2} |\Phi\rangle|0\rangle_{B_2} =  \alpha |E\rangle |1\rangle_{B_1}  |1\rangle_{B_2} + \beta |F\rangle |0\rangle_{B_1} |0\rangle_{B_2}.
\end{eqnarray}
After that we apply  QCM $W^{(3)}_{B_1B_2}$, defined in Eq.(\ref{CM}), to $|\Phi_1\rangle$ { thus selecting the term with the state $|1\rangle_{B_2}$  from the superposition present in Eq.(\ref{SPhi1}) and obtain the following normalized resulting state, according to Eq.(\ref{tMf2}):}
\begin{eqnarray}\label{E2}
|\Phi_2\rangle = W^{(3)}_{B_1B_2} |\Phi_1\rangle =  {  \alpha |E\rangle \Big(M_{B_2} |1\rangle_{B_1}  |1\rangle_{B_2}\Big) + \beta |F\rangle |0\rangle_{B_1} |0\rangle_{B_2}   =}
{ \frac{\alpha}{|\alpha|}}  |E\rangle |1\rangle_{B_1} {  |1\rangle_{B_2}}.
\end{eqnarray}
In this way,   the problem of small success probability  in the matrix-manipulation algorithms  can be removed.
 The scheme for replacing the usual measurement by  QCM is shown in  Fig.\ref{Fig:CM}.

\begin{figure*}[!]
\centering
{\includegraphics[width=4in, angle=0]{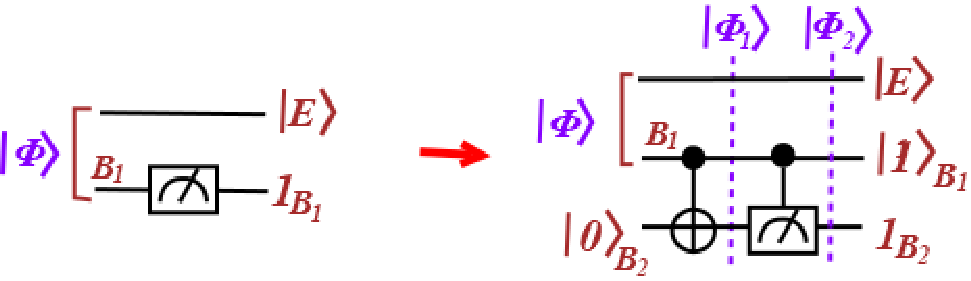}}
\caption{Replacement of the ordinary measurement of the ancilla state (left circuit) with the  subroutine of  QCM (right circuit).  The state $|\Phi\rangle$ and the intermediate states $|\Phi_j\rangle$, $j=1,2$, are determined, respectively, in Eqs.(\ref{EF}), (\ref{SPhi1}) and (\ref{E2}).} 
\label{Fig:CM}
\end{figure*}

\subsection{Multiple random choice { in superposition} state}

We consider the superposition state
\begin{eqnarray}\label{Phi00}
|\Phi_0\rangle =\sum_{j=1}^Q \alpha_j |E_j\rangle +\beta |F\rangle , \;\; \langle E_j|F\rangle =0,\;\;
\langle E_j|E_k\rangle =\delta_{jk}, \;\;\sum_{j=1}^Q|\alpha_j|^2+|\beta|^2=1,
\end{eqnarray}
where $|E_j\rangle$, { $j=1,\dots, Q$,}  and $|F\rangle$ are orthonormal  states of some quantum system, and  the states  $|E_j\rangle$ contain the useful information.
We  label the states $|E_j\rangle$ and  $|F\rangle$ using $(\lceil \log (Q+1)\rceil+1)$-qubit ancilla $B$:
\begin{eqnarray}\label{Phi01}
|\Phi_0\rangle \to |\Phi_1\rangle = \sum_{j=1}^Q \alpha_j |E_j\rangle|1\rangle_{B_1} |j\rangle_{B'} +
\beta |F\rangle|0\rangle_{B_1}|0\rangle_{B'} ,
\end{eqnarray}
{ where $B_1$   is the first qubit of the ancilla $B$} and $B'$ collects all other qubits of $B$.  { We assume that} such labeling can be  done  using the particular structures of the states   $|E_j\rangle$, see, for instance, Refs. \cite{ZQKW_2024,ZBQKW_arXive2024}.

Now we introduce  QCM, { similar to that given in (\ref{CM}),}
\begin{eqnarray}\label{CM22}
W_{B} =  |1\rangle_{B_1}\,{ _{B_1}\langle 1|}\otimes M_{B'} +|0\rangle_{B_1}\,{ _{B_1}\langle 0|} \otimes I_{B'},
\end{eqnarray}
where $M_{B'}$ is the measurement operator applied to each qubit of $B'$.  { Applying  QCM $W_{B}$ to $|\Phi_1\rangle$ we fix  $B'$ in one of the  states $|j\rangle_{B'}$, $j\neq 0$,  with the appropriate probability $p_j$, and this state is attached to the state $|E_j\rangle$. Thus, the garbage (the second term in Eq.(\ref{Phi01})) disappears from the result, which is reflected in the following formula:}
\begin{eqnarray}\label{chi22}
|\Phi_2\rangle =W_{B} |\Phi_1\rangle=M_{B'} \sum_{j=1}^Q \alpha_j  |E_j\rangle |1\rangle_{B_1}  |j\rangle_{B'}  { +
\beta |F\rangle|0\rangle_{B_1}|0\rangle_{B'}}={    \sum_{j=1}^Q p_j  |E_j\rangle  |1\rangle_{B_1}   { |j\rangle_{B'} .}}
\end{eqnarray}
In other words, the result of measurement $M_{B'}$ yields the state $ |E_j\rangle  |1\rangle_{B_1}$ with the probability
$p_j=|\alpha_j|^2/\sum_{j=1}^Q|\alpha_j|^2$, which does not depend on the probability amplitude $\beta$.

\section{Extensions of matrix encoding technique}
\label{Section:extension}
{     Both extensions of matrix-manipulation algorithms represented in Secs.\ref{Section:HC} and  \ref{Section:norm} are based on the modification of matrix encoding into the initial state of a quantum system. Therefore, we  first recall the matrix encoding into the initial state of the matrix-multiplication algorithm \cite{ZQKW_2024,ZBQKW_arXive2024} (the initial sate for the matrix-inversion algorithm \cite{ZBQKW_arXive2024} can be modified similarly). }

The matrix  encoding implemented   in the algorithm for matrix multiplication  uses two $n$-qubit quantum subsystems $R$ and $C$ whose basis states label, respectively,  the rows and columns associated  with each element $a_{jk}$ (complex in general) of the $N\times N$ matrix
$A=\{a_{jk}: j,k=0,\dots,N-1\}$, $N=2^n$. Thus,  the matrix $A$ is encoded as follows:
\begin{eqnarray}\label{inst}
|\Psi_A\rangle &=&
 \sum_{j=0}^{N-1} \sum_{k=0}^{N-1}a_{jk}   |j\rangle_{R} |k\rangle_{C}
 \end{eqnarray}
 with the following normalization condition for the pure state of a quantum system:
 \begin{eqnarray}\label{constr0}
 \sum_{j=0}^{N-1} \sum_{k=0}^{N-1}  |a_{jk}|^2=1.
\end{eqnarray}
We shall note that the  initial state   preparation in Eq. (\ref{inst}) is a special task by itself and might be time-consuming operation.  The problem of an {\it arbitrary} initial state encoding was considered in Ref.
\cite{ZQW}.
It is shown that the  depth of such algorithm is $O(N^2 \log N)$ taking into account post-selection. Without post-selection the depth is $O(N \log N)$. The feature of that algorithm of matrix encoding is that it does not require classical support for calculating the parameters of encoding unitary transformations. Although the depth of the state encoding algorithm seems to be large, the encoding expanses can be covered if the set of successive matrix operations is assumed over  the given  set of encoded matrices (input data).  We also refer to several other state encoding algorithms  \cite{VMS,PFTV,STYYZ,YZ} that require certain classical support for calculating the parameters of encoding operators. 

Below, we introduce two extensions to the algorithms of matrix manipulations.
The first of them is aimed on effective implementing the operations with complex matrices, Sec.\ref{Section:HC}.  The second one allows to weaken the  constraint,  Eq.(\ref{constr0}), imposed on the matrix elements by the normalization condition holding for  pure quantum states, Sec.\ref{Section:norm}. We deal with the square $N\times N$ matrices for convenience. If the original matrix is rectangular we turn it to the square matrix by appending the appropriate number of either  zero rows or zero columns.
In Sec. \ref{Section:newencoding}, both above extensions are combined in the single circuit.
Then, in Sec.\ref{Section:mult}, both these extensions will be implemented in the matrix-multiplication algorithm. Therein, we also compare the { results of} applying  the usual measurement and  QCM.

\subsection{Hermitian conjugate}
\label{Section:HC}

 The complex elements of the $N\times N$ matrix $A=\{a_{jk}: j,k=0,\dots N-1\}$ can be written  as
\begin{eqnarray}
\label{a}
a_{jk}=a_{jk0}+ i a_{jk1}.
\end{eqnarray}
 We again involve two mentioned above  $n$-qubit subsystems $R$ nd $C$ and add the third one-qubit subsystem $M$ whose state   is $|0\rangle_M$ for the real part of $a_{jk}$ and $|1\rangle_M$ for the imaginary part.
Then  the elements $a_{jk}$ of the matrix $A$ can be encoded into the superposition state of three subsystems $R$, $C$ and  $M$ as follows:
\begin{eqnarray}\label{inst0}
|\Psi_A\rangle= \sum_{j=0}^{N-1}\sum_{k=0}^{N-1} \sum_{m=0}^1 a_{jkm} |j\rangle_R |k\rangle_C |m\rangle_M.
\end{eqnarray}
With such pure state,
the  Hermitian conjugation of $A$ can be realized by the following operator:
\begin{eqnarray}\label{SWAPM}
W=SWAP_{RC} \sigma^{(z)}_M,
\end{eqnarray}
where $SWAP_{RC}$ exchanges states of the subsystems $R$ and $C$ (in other words, it performs transposition of $A$), and  $\sigma^{(z)}_M$ is the $\sigma^{(z)}$-operator applied to $M$. This operator changes the sign of  the  excited state $|1\rangle_M$ and thus  performs the complex conjugation of $A$.
Therefore,
the matrix encoded into the state $W |\Psi_A\rangle$ is the Hermitian conjugate of $A$,
\begin{eqnarray}\label{tPhiA}
|\tilde \Psi_A\rangle = W |\Psi_A\rangle = \sum_{j=0}^{N-1}\sum_{k=0}^{N-1}  \sum_{m=0}^1 a_{kjm} (-1)^m |j\rangle_R |k\rangle_C |m\rangle_M.
\end{eqnarray}
The Hermitian conjugation algorithm can be used as a subroutine in other algorithms, as shown in Sec.\ref{Section:manipulations}. The circuit for the Hermitian  conjugation is presented in Fig.\ref{Fig:HC}.

\begin{figure}[ht]
    \includegraphics[width=0.15\textwidth]{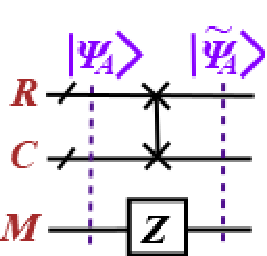}
    \caption{The circuit for the Hermitian conjugation, $Z\equiv \sigma^{(z)}$. The states $|\Psi_A\rangle$ and $|\tilde \Psi_A\rangle$ are determined, respectively, in Eqs. (\ref{inst0}) and  (\ref{tPhiA}).}
\label{Fig:HC}
\end{figure}
We shall notice that using the new encoding of the complex matrix in  the superposition state of the triple system $R\otimes C\otimes M$ requires appropriate modifications for the  multiplication algorithm which is considered in Sec.\ref{Section:mult}.
In addition, since the algorithms for determinant calculation and matrix inversion \cite{ZBQKW_arXive2024} use the row-wise matrix encoding, {each row might  require its own additional  one-qubit subsystem to label the real and imaginary parts of its matrix elements, i.e., at most $N$ one-qubit  subsystems $M_i$, $0\le i \le N-1$, are required for those algorithms.}

\subsection{Weakened normalization constraint for matrix encoding}
\label{Section:norm}

Eq.(\ref{constr0}) represents a constraint imposed  on the { modulus} of the matrix elements. However, we can weaken this constraint reducing it to the inequality
\begin{eqnarray}\label{constr1}
 \sum_{j=0}^{N-1} \sum_{k=0}^{N-1}  |a_{jk}|^2 \le  1.
\end{eqnarray}
This can be done if 
we
 introduce, along with the subsystems $R$ and $C$ enumerating rows and columns of the $2^n \times 2^n$ complex matrix $A$,  the one-qubit subsystem $K$ and replace the initial  state given in  Eq.(\ref{inst}) with the following one:
 \begin{eqnarray}\label{instm}
|\Psi_A\rangle &=&
b  |0\rangle_{R} |0\rangle_{C}|0\rangle_K+
 \sum_{j=0}^{N-1} \sum_{k=0}^{N-1} a_{jk}   |j\rangle_{R} |k\rangle_{C}|1\rangle_K
 \end{eqnarray}
 with the normalization
 \begin{eqnarray}\label{constr2}
|b|^2+ \sum_{j=0}^{N-1} \sum_{k=0}^{N-1}  |a_{jk}|^2=1.
\end{eqnarray}
Eq.(\ref{constr2}) yields  inequality (\ref{constr1}). Thus, replacing Eq.(\ref{constr0}) with inequality (\ref{constr1}) becomes possible due to introducing the additional probability amplitude $b$ into the initial  state,
Eq.(\ref{instm}). {  Again, to weaken the normalization constraint in the row-wise encoding \cite{ZBQKW_arXive2024}, an additional qubit might be required for encoding each row.  }

 \subsection{Circuit for new encoding  of initial state}
 \label{Section:newencoding}
Now we combine the modifications of  the state encoding presented in Eqs.(\ref{inst0}) and (\ref{instm}) and encode the elements $ a_{ij}$ decomposed into the real and imaginary parts,  shown in  Eq.(\ref{a}),  as  follows:
\begin{eqnarray}\label{inst2}
|\Psi_A\rangle &=&( b_{0}|0\rangle_{M} +b_{1}|1\rangle_{M})  |0\rangle_{R} |0\rangle_{C}|0\rangle_{K}
+\\\nonumber
&&
 \sum_{j=0}^{N-1}   \sum_{k=0}^{N-1}  ( a_{jk0}|0\rangle_{M} + a_{jk1}|1\rangle_{M})   |j\rangle_{R} |k\rangle_{C}|1\rangle_{K},\;\; b=b_0 + i b_1,\\\label{norm}
&&|b|^2+\sum_{j=0}^{N-1}   \sum_{k=0}^{N-1} |a_{jk}|^2=1.
\end{eqnarray}
Replacement of matrix encoding  via Eq. (\ref{inst}) with encoding via Eq.(\ref{inst2}) is shown in Fig.\ref{Fig:mod}.
\begin{figure}[ht]
    \includegraphics[width=0.4\textwidth]{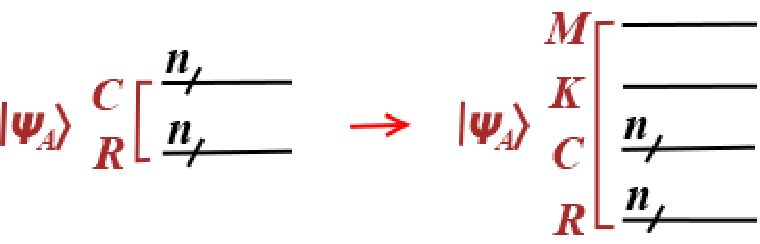}
    \caption{Replacement of the matrix encoding via Eq.(\ref{inst}) with the matrix encoding via Eq.(\ref{inst2}) for organizing the effective manipulations with complex matrices and weakening the constraint on the matrix elements. The auxiliary one-qubit subsystems $M$ and $K$ are added to the $n$-qubit subsystems $C$ and $R$ whose basis states enumerate, respectively, the columns and rows of the encoded $2^n\times 2^n$ matrix with the elements $a_{jk}$.}
\label{Fig:mod}
\end{figure}

\section{Matrix multiplication}
\label{Section:mult}
In this section  we propose the matrix multiplication algorithm \cite{ZQKW_2024,ZBQKW_arXive2024} that takes into account the extensions  discussed in Secs. \ref{Section:extension}, as well as implements  QCM described in Sec.\ref{Section:contrmeas}.

Proposed modification is based on  passing from encoding via Eq.(\ref{inst}) (used  in \cite{ZQKW_2024} )  to encoding via Eq.(\ref{inst2}) for the  square  $N\times N$   matrices  $A^{(1)}$ and $A^{(2)}$ ($A^{(l)}=\{a^{(l)}_{jk}: j,k =0,\dots,N-1\}$, $l=1,2$)
 according to Sec.\ref{Section:extension}.

%
We first introduce four $n$-qubit   registers   $R^{(l)}$ and $C^{(l)}$ , $l=1,2$, enumerating, respectively, the
 rows and columns of the matrices $A^{(l)}$, $l=1,2$. Then, we introduce four one-qubit subsystems $M^{(l)}$ and $K^{(l)}$,
 $l=1,2$.  The subsystems $M^{(l)}$, $l=1,2$, are used to label the real and imaginary parts of the matrix elements of the $l$th
 matrix ($a^{(l)}_{jk} = a^{(l)}_{jk0} + i a^{(l)}_{jk1}$), and also to label  the real and imaginary parts of the parameters
 $b^{(l)}$   ($b^{(l)} = b^{(l)}_{0} + i b^{(l)}_{1}$). The subsystems $K^{(l)}$, $l=1,2$,  are needed to
 weaken the constraints imposed  on  the elements  $a^{(l)}_{jk}$  by the normalization condition for a pure quantum state.
Thus, the pure states encoding the elements of matrices $A^{(l)}$, $l=1,2$, are in the form of Eq.(\ref{inst2}) with normalization given in  Eq.(\ref{norm}):
\begin{eqnarray}\label{inst23}
|\Psi^{(l)}\rangle &=&( b^{(l)}_{0}|0\rangle_{M^{(l)}} +b^{(l)}_{1}|1\rangle_{M^{(l)}})  |0\rangle_{R^{(l)}} |0\rangle_{C^{(l)}}|0\rangle_{K^{(l)}}
+ \\\nonumber
&&\sum_{j,k=0}^{N-1}  (a^{(l)}_{jk0}|0\rangle_{M^{(l)}} +a^{(l)}_{jk1}|1\rangle_{M^{(l)}})   |j\rangle_{R^{(l)}} |k\rangle_{C^{(l)}}|1\rangle_{K^{(l)}},\\\label{norm2}
&&|b^{(l)}|^2+\sum_{j,k} |a^{(l)}_{jk}|^2=1,\;\;l=1,2,
\end{eqnarray}
where the constant parameters $b^{(l)}$ are introduced  in accordance with Sec.\ref{Section:norm}.
The initial state of the whole system reads
\begin{eqnarray}\label{Phi0}
&&|\Phi_0\rangle =  |\Psi^{(2)}\rangle \otimes |\Psi^{(1)}\rangle=
\\\nonumber &&
\sum_{m_1,m_2=0}^1
b^{(1)}_{m_1} b^{(2)}_{m_2} |m_1\rangle_{M^{(1)}}|m_2\rangle_{M^{(2)}}  |0\rangle_{R^{(1)}} |0\rangle_{C^{(1)}} |0\rangle_{R^{(2)}} |0\rangle_{C^{(2)}} |0\rangle_{K^{(1)}}|0\rangle_{K^{(2)}}+\\\nonumber
&&
   \sum_{{j_1,k_1,}\atop{j_2,k_2=0}}^{N-1}\sum_{m_1,m_2=0}^1
\left(a^{(1)}_{j_1k_1m_1}a^{(2)}_{j_2k_2m_2}|m_1\rangle_{M^{(1)}}|m_2\rangle_{M^{(2)}} \right)\\\nonumber
&&\times |j_1\rangle_{R_1} |k_1\rangle_{C^{(1)}} |j_2\rangle_{R^{(2)}} |k_2\rangle_{C^{(2)}}|1\rangle_{K^{(1)}}|1\rangle_{K^{(2)}}+
|g_0\rangle,
\end{eqnarray}
where we select the terms with the states $|j\rangle_{K^{(1)}}|j\rangle_{K^{(2)}}$, $j=0,1$,  which are  needed below, while the terms with the states
$|j\rangle_{K^{(1)}}|k\rangle_{K^{(2)}}$, $j\neq k$,  are  collected in the garbage $|g_0\rangle$.

From the superposition state $|\Phi_0\rangle$, we have to select the terms with the states
$|k\rangle_{C^{(1)}} |k\rangle_{R^{(2)}}$, $k=0,\dots,N-1$, i.e.,  $k_1=j_2$ in the second part in the right-hand side of Eq.(\ref{Phi0}).
For that purpose we apply the C-NOTs $W_j$ to the $j$th qubits of $C^{(1)}$ and $R^{(2)}$,
\begin{eqnarray}
W_{j} &=& |1\rangle_{C^{(1)}_j}\; _{C^{(1)}_j}\langle 1| \otimes \sigma^{(x)}_{R^{(2)}_j} +
 |0\rangle_{C^{(1)}_j}\; _{C^{(1)}_j}\langle 0| \otimes I_{R^{(2)}_j},\;\;j=1,\dots,n.
\end{eqnarray}
where  $\sigma^{(x)}_{R^{(2)}_j}$ and $I_{R^{(2)}_j}$  are, respectively,  the  $\sigma^{(x)}$-operator and identity operator applied to the $j$th qubit of the subsystem $R^{(2)}$.
{We have chosen the qubits of   $C^{(1)}$ as controlling ones. }
All operators $W_j$, $j=1,\dots, n$, commute with each other.
Applying the operator $W^{(0)}_{C^{(1)}R^{(2)}}$,
\begin{eqnarray}\label{W1m}
W^{(0)}_{C^{(1)}R^{(2)}}=\prod_{j=1}^n W_j,
\end{eqnarray}
   to $|\Phi_0\rangle$ we obtain:
\begin{eqnarray}\label{Phi11}
&&
|\Phi_1\rangle =W^{(0)}_{C^{(1)}R^{(2)}} |\Phi_0\rangle
=\\\nonumber
&&
\sum_{m_1,m_2=0}^1
b^{(1)}_{m_1} b^{(2)}_{m_2} |m_1\rangle_{M^{(1)}}|m_2\rangle_{M^{(2)}}  |0\rangle_{R^{(1)}} |0\rangle_{C^{(1)}} |0\rangle_{R^{(2)}} |0\rangle_{C^{(2)}}  |0\rangle_{K^{(1)}}|0\rangle_{K^{(2)}}+\\\nonumber
&&
\sum_{{j_1,j,}\atop{k_2=0}}^{N-1}\sum_{m_1,m_2=0}^1
\left(a^{(1)}_{j_1jm_1}a^{(2)}_{jk_2m_2}|m_1\rangle_{M^{(1)}}|m_2\rangle_{M^{(2)}} \right)\\\nonumber
&&\times
  |j_1\rangle_{R^{(1)}} |j\rangle_{C^{(1)}} |0\rangle_{R^{(2)}} |k_2\rangle_{C^{(2)}}
|1\rangle_{K^{(1)}}|1\rangle_{K^{(2)}}+|g_1\rangle. 
\end{eqnarray}
The first and second  parts in  the  right-hand side of Eq.(\ref{Phi11}) collect the terms  with  the state $|0\rangle_{R_2} |j\rangle_{K^{(1)}}|j\rangle_{K^{(2)}}$, $j=0,1$,
(the other terms are  garbage by definition, they are collected in
$ |g_1\rangle$ above). We note that the first part in the right-hand side of Eq.(\ref{Phi11})  remains unchanged after applying the operator $W^{(0)}_{C^{(1)}R^{(2)}}$.
We also note that the operator $W^{(0)}_{C^{(1)}R^{(2)}}$   transforms $|g_0\rangle$ as well.
However, we don't need details of that transformation. The similar  statement  can be referred  to the operators $ W^{(1)}_{C^{(1)}}$, $W^{(2)}_{M^{(1)}M^{(2)}K^{(1)}K^{(2)}}$ and  $W^{(3)}_{C^{(1)} R^{(2)} M^{(2)}K^{(2)}B}$ below.

Next,  we apply the Hadamard transformations
\begin{eqnarray}
\label{W2m}
W^{(1)}_{C^{(1)}}=H^{\otimes  n}
\end{eqnarray}
to the qubits of  $C^{(1)}$  and select the terms with $|0\rangle_{C^{(1)}} |0\rangle_{R^{(2)}}|j\rangle_{K^{(1)}}|j\rangle_{K^{(2)}}$, $j=0,1$, putting others to the garbage $|g_2\rangle$:
\begin{eqnarray}\label{Phi32}
&&
|\Phi_2\rangle =W^{(1)}_{C^{(1)}}  |\Phi_1\rangle
 =\\\nonumber
&&\frac{1}{2^{n/2}}
\sum_{m_1,m_2=0}^1
b^{(1)}_{m_1} b^{(2)}_{m_2} |m_1\rangle_{M^{(1)}}|m_2\rangle_{M^{(2)}}  |0\rangle_{R^{(1)}} |0\rangle_{C^{(1)}} |0\rangle_{R^{(2)}} |0\rangle_{C^{(2)}}  |0\rangle_{K^{(1)}}|0\rangle_{K^{(2)}}\\\nonumber
&&+\frac{1}{2^{n/2}}
\sum_{j_1,k_2=0}^{N-1}  \sum_{ j=0}^{N-1}  \sum_{m_1,m_2=0}^1
\left(a^{(1)}_{j_1jm_1}a^{(2)}_{jk_2m_2}|m_1\rangle_{M^{(1)}}|m_2\rangle_{M^{(2)}} \right)\\\nonumber
&&\times
  |j_1\rangle_{R^{(1)}} |0\rangle_{C^{(1)}} |0\rangle_{R^{(2)}} |k_2\rangle_{C^{(2)}}
|1\rangle_{K^{(1)}}|1\rangle_{K^{(2)}}+
|g_2\rangle. 
\end{eqnarray}
Now, to complete multiplication,  we have to organize proper combinations of products  of real and imaginary parts of matrices $A^{(i)}$, $i=1,2$, to form the real and imaginary parts of the result. To this end we  introduce
the following controlled operator:
\begin{eqnarray}\label{M1M1}
W_{M^{(1)}M^{(2)}} = |1\rangle_{M^{(2)}} \, _{M^{(2)}}\langle 1| \otimes \sigma^{(x)}_{M^{(1)}}\sigma^{(z)}_{M^{(1)}}  +
 |0\rangle_{M^{(2)}} \, _{M^{(2)}}\langle 0| \otimes I_{M^{(1)}}.
\end{eqnarray}
We also introduce the controlled operator
\begin{eqnarray}\label{K1K2}
W_{K^{(1)}K^{(2)}} = |1\rangle_{K^{(1)}} \, _{K^{(1)}}\langle 1 | \otimes \sigma^{(x)}_{K^{(2)}}  + |0\rangle_{K^{(1)}} \, _{K^{(1)}}\langle 0 | \otimes I_{K^{(2)}}
\end{eqnarray}
which labels all needed terms by the state  $|0\rangle_{K^{(2)}}$.
We include the operators $ W_{M^{(1)}M^{(2)}}$ and $W_{K^{(1)}K^{(2)}}$ into   the operator
\begin{eqnarray}\label{W1}
W^{(2)}_{M^{(1)}M^{(2)}K^{(1)}K^{(2)}}=W_{K^{(1)}K^{(2)}}H_{M^{(2)}}W_{M^{(1)}M^{(2)}},
\end{eqnarray}
 where $H_{M^{(2)}}$ is the  Hadamard operator applied to the single  qubit of $M^{(2)}$, and apply
 $W^{(2)}_{M^{(1)}M^{(2)}K^{(1)}K^{(2)}}$  to $|\Phi_2\rangle $, thus obtaining
\begin{eqnarray}\label{Phi2}
&&
|\Phi_3\rangle =W^{(2)}_{M^{(1)}M^{(2)}K^{(1)}K^{(2)}}  |\Phi_2\rangle
 =\\\nonumber
&&\frac{1}{2^{(n+1)/2}}\Big(
(b^{(1)}_{0} b^{(2)}_{0}-b^{(1)}_{1} b^{(2)}_{1})|0\rangle_{M^{(1)}} +
(b^{(1)}_{0} b^{(2)}_{1}+b^{(1)}_{1} b^{(2)}_{0})|1\rangle_{M^{(1)}}\Big) \\\nonumber
&&\times
 |0\rangle_{M^{(2)}} |0\rangle_{R^{(1)}} |0\rangle_{C^{(1)}} |0\rangle_{R^{(2)}} |0\rangle_{C^{(2)}}|0\rangle_{K^{(1)}}|0\rangle_{K^{(2)}} \\\nonumber
&&
+\frac{1}{2^{(n+1)/2}} \sum_{j_1,k_2=0}^{N-1} \sum_{j=0}^{N-1}   \left((a^{(1)}_{j_1j0}a^{(2)}_{j k_20}-
a^{(1)}_{j_1 j 1}a^{(2)}_{j k_2 1})|0\rangle_{M^{(1)}}
\right.\\\nonumber
&&\left.+
(a^{(1)}_{j_1j0}a^{(2)}_{jk_21}+a^{(1)}_{j_1j1}a^{(2)}_{jk_20})|1\rangle_{M^{(1)}} \right) \\\nonumber
&&\times
|0\rangle_{M^{(2)}}   |j_1\rangle_{R^{(1)}} |0\rangle_{C^{(1)}} |0\rangle_{R^{(2)}} |k_2\rangle_{C^{(2)}}|1\rangle_{K^{(1)}}|0\rangle_{K^{(2)}}  +
|g_3\rangle.
\end{eqnarray}
Here the first term plays an auxiliary role similar to the first term in Eq.(\ref{inst23}), while the second  term contains the desired matrix product. All extra terms are collected in the garbage
$|g_3\rangle$.
We emphasize that the operator
$W^{(2)}_{M^{(1)}M^{(2)}K^{(1)}K^{(2)}}$ is a novelty in comparison with multiplication algorithm in
 \cite{ZBQKW_arXive2024}.

Next, to label and remove the  garbage, we prepare the two-qubit ancillae $ B$   in the ground state $|0\rangle_{B}$, introduce the
projector
\begin{eqnarray}
P_{C^{(1)}R^{(2)}M^{(2)}K^{(2)}} = |0\rangle_{C^{(1)}}|0\rangle_{R^{(2)}} |0\rangle_{M^{(2)}}|0\rangle_{K^{(2)}}
 \;  _{C^{(1)}}\langle 0|   _{R^{(2)}}\langle 0|   _{M^{(2)}}\langle 0|   _{K^{(2)}}\langle 0|
\end{eqnarray}
and the  controlled operator $W^{(3)}_{C^{(1)} R^{(2)} M^{(2)}K^{(2)} B}$,
\begin{eqnarray}
\label{W3m}
&&
W^{(3)}_{C^{(1)} R^{(2)} M^{(2)}K^{(2)}B} = P_{C^{(1)}R^{(2)}M^{(2)}K^{(2)}} \otimes \sigma^{(x)}_{B_1} \sigma^{(x)}_{B_2}\\\nonumber
&& + (I_{C^{(1)}R^{(2)}M^{(2)}K^{(2)}}- P_{C^{(1)}R^{(2)}M^{(2)}K^{(2)}})\otimes I_{B},
\end{eqnarray}
 of the depth $O(n)$ with $2(n+1)$-qubit control register. Here $B_i$, $i=1,2$, means the $i$th qubit of the ancilla $B$.
Applying this operator  to $|\Phi_3\rangle |0\rangle_{B}$ we obtain
\begin{eqnarray}\label{Phi4}
&&
|\Phi_4\rangle = W^{(3)}_{C^{(1)} R^{(2)} M^{(2)}K^{(2)} B}|\Phi_3\rangle |0\rangle_{ B}  =
\\\nonumber
&&
\frac{1}{2^{ (n+1)/2}}\Big(
(b^{(1)}_{0} b^{(2)}_{0}-b^{(1)}_{1} b^{(2)}_{1})|0\rangle_{M^{(1)}} +
(b^{(1)}_{0} b^{(2)}_{1}+b^{(1)}_{1} b^{(2)}_{0})|1\rangle_{M^{(1)}}\Big) \\\nonumber
&&\times |0\rangle_{M^{(2)}}
 |0\rangle_{R^{(1)}} |0\rangle_{C^{(1)}} |0\rangle_{R^{(2)}} |0\rangle_{C^{(2)}}|0\rangle_{K^{(1)}}|0\rangle_{K^{(2)}} |1\rangle_{ B_1}  |1\rangle_{B_2} +\\\nonumber
&&
\frac{1}{2^{ (n+1)/2}}\sum_{{j_1,k_2,}\atop{j=0}}^{N-1}
\Big((a^{(1)}_{j_1j0}a^{(2)}_{jk_20}-a^{(1)}_{j_1j1}a^{(2)}_{jk_21})|0\rangle_{M^{(1)}} +(a^{(1)}_{j_1j0}a^{(2)}_{jk_21}+a^{(1)}_{j_1j1}a^{(2)}_{jk_20})|1\rangle_{M^{(1)}} \Big)
\\\nonumber
&&\times
|0\rangle_{M^{(2)}}   |j_1\rangle_{R^{(1)}} |0\rangle_{C^{(1)}} |0\rangle_{R^{(2)}} |k_2\rangle_{C^{(2)}}|1\rangle_{K^{(1)}}|0\rangle_{K^{(2)}}\ |1\rangle_{B_1}|1\rangle_{B_2}+  |g_3\rangle 
 |0\rangle_{B_1} |0\rangle_{B_2}.
\end{eqnarray}
{    Now we have to remove the garbage marked by the state  $|0\rangle_{B_1} |0\rangle_{B_2}$.  { Before implementing  QCM described in Sec.\ref{Section:CM}, we apply the  most straightforward way of garbage removal that}  is the usual  measurement of the state of $B_2$  with the  desired output $|1\rangle_{B_2}$.  { Of course, this way requires multiple running of the algorithm because the single run with the subsequent measurement might yield undesired result $|0\rangle_{B_2}$ that must be disregarded.}
But once the desired output is obtained, the quantum state of the remaining  system becomes}
\begin{eqnarray}\label{Phi6}
&&
|\Phi_5\rangle =
|\Psi_{out}\rangle\,|0\rangle_{M_2} |0\rangle_{C_1} |0\rangle_{R_2}|0\rangle_{K_2}|1\rangle_{B_1}  ,
\\\label{PsiOut}
&&
|\Psi_{out}\rangle =G^{{-1}}\Big(  \Big(
(b^{(1)}_{0} b^{(2)}_{0}-b^{(1)}_{1} b^{(2)}_{1})|0\rangle_{M^{(1)}} +
(b^{(1)}_{0} b^{(2)}_{1}+b^{(1)}_{1} b^{(2)}_{0})|1\rangle_{M^{(1)}}\Big)\\\nonumber
&&\times |0\rangle_{R^{(1)}} |0\rangle_{C^{(2)}}   |0\rangle_{K_1}     +
\\\nonumber
&&
    \sum_{{j_1,k_2,}\atop{j=0}}^{N-1}
\left((a^{(1)}_{j_1j0}a^{(2)}_{jk_20}-a^{(1)}_{j_1j1}a^{(2)}_{jk_21})|0\rangle_{M^{(1)}} +(a^{(1)}_{j_1j0}a^{(2)}_{jk_21}
+a^{(1)}_{j_1j1}a^{(2)}_{jk_20})|1\rangle_{M^{(1)}} \right)
 \\\nonumber
&&\times
 |j_1\rangle_{R^{(1)}} |k_2\rangle_{C^{(2)}} |1\rangle_{K^{(1)}}\Big)=\\\nonumber
&&
G^{-1}\left( (\hat b_0 |0\rangle_{M^{(1)}}    +\hat b_1|1\rangle_{M^{(1)}} )  |0\rangle_{R^{(1)}} |0\rangle_{C^{(2)}}  |0\rangle_{K^{(1)}} \right. \\\nonumber
&&\left.+
\sum_{j,k=0}^{N-1}(\hat a_{jk0}|0\rangle_{M^{(1)}} +\hat a_{jk1}|1\rangle_{M^{(1)}})| j\rangle_{R^{(1)}} |k\rangle_{C^{(2)}}|1\rangle_{K^{(1)}} \right),
\end{eqnarray}
where  $\displaystyle G=\left(|b^{(1)} b^{(2)}|^2+\sum_{j,k=0}^{N-1} \Big| \sum_{l=0}^{N-1} a^{(1)}_{j l}a^{(2)}_{lk}\Big|^2 \right)^{1/2}$ is the normalization factor,
\begin{eqnarray}
&&
\hat b_0 = b^{(1)}_{0} b^{(2)}_{0}-b^{(1)}_{1} b^{(2)}_{1}, \;\;
 \hat b_1=b^{(1)}_{0} b^{(2)}_{1}+b^{(1)}_{1} b^{(2)}_{0},\\\nonumber
&&
\hat a_{jk0} =  \sum_{l=0}^{N-1}
(a^{(1)}_{jl0}a^{(2)}_{lk0}-a^{(1)}_{jl1}a^{(2)}_{lk1}),\;\;
\hat a_{jk1} = \sum_{l=0}^{N-1}(a^{(1)}_{jl0}a^{(2)}_{lk1}+a^{(1)}_{jl1}a^{(2)}_{lk0}),
\end{eqnarray}
so that $\hat A= A^{(1)} A^{(2)} = \{\hat a_{jk}: j,k=0,\dots,N-1 \}$, $\hat a_{jk} = \hat a_{jk0} + i  \hat a_{jk1}$,
$\hat b= \hat b_0 + i \hat b_1$, where $ \hat a_{jkm}$, $\hat b_m$, $m=0,1$, are real numbers. We note that the first term in the state $|\Psi_{out}\rangle$ given in Eq.(\ref{PsiOut}) plays  the same role as the first term in the expressions for $|\Psi^{(l)}\rangle$, given in  Eq.(\ref{inst23}). It weakens the constraint on the normalization of the elements of the resulting  matrix $\hat A$.

{The weak point is that the probability of measuring the desired state $|1\rangle_{B_2}$ is small,
$p=G^2/2^{n+1}$, and exponentially decreases with an increase in $n$. Therefore, to obtain the desired result $|1\rangle_{B_2}$ we have to run the algorithm $O(2^{n+1}/G^2)$ times.
However, we can replace the usual  measurement by  QCM, i.e.,}
let us  perform  QCM of the ancilla qubit $B_2$ via the operator {  (see also Eq.(\ref{CM}) in Sec.\ref{Section:CM})}
\begin{eqnarray}
W^{(4)}_{B} = |1\rangle_{B_1}\,_{B_1}\langle 1| \otimes M_{ B_2}  + |0\rangle_{B_1}\,_{B_1}\langle 0| \otimes I_{B_2},
\end{eqnarray}
where $M_{B_2}$ is the measurement operator applied to the qubit $B_2$.
{Then, {  applying $W^{(4)}_{B}$ to $|\Phi_4\rangle$ , we select the term with the state $|1\rangle_{B_2}$ from this superposition, thus removing the garbage and obtaining:}
\begin{eqnarray}\label{Phi62}
&&
|\Phi_5\rangle = W^{(4)}_{B} |\Phi_4\rangle=
|\Psi_{out}\rangle\,|0\rangle_{M_2} |0\rangle_{C_1} |0\rangle_{R_2}|0\rangle_{K_2}|1\rangle_{B_1} { |1\rangle_{B_2}} ,
\end{eqnarray}
where $|\Psi_{out}\rangle$ is determined in Eq.(\ref{PsiOut}). In this case, only the  single run of the algorithm is enough to get the desired state $|\Psi_{out}\rangle$.}


\begin{figure*}[!]
\centering
    \centerline{\includegraphics[width=0.9\textwidth]{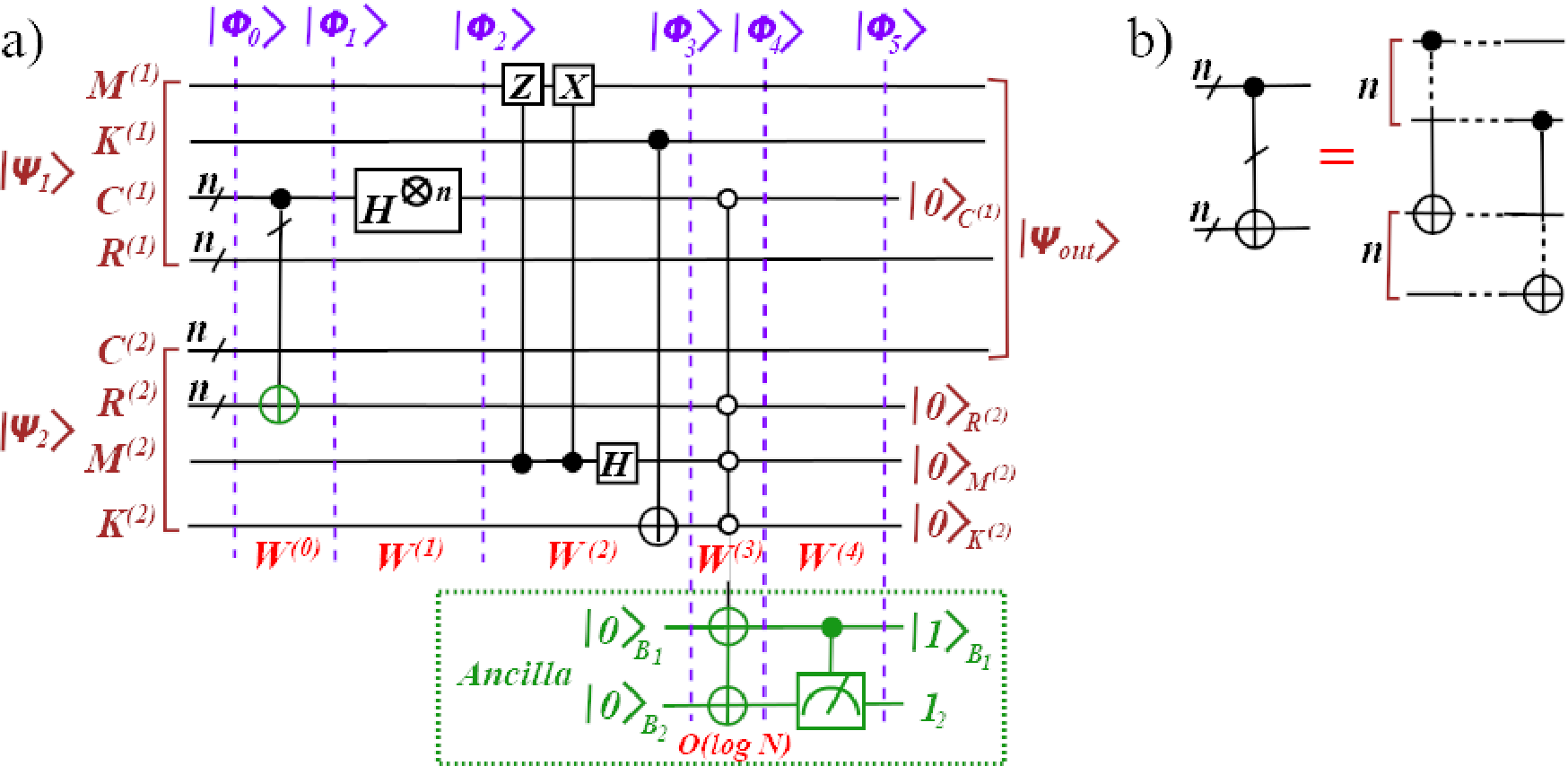}}
\caption{(a) The circuit for the matrix multiplication algorithm; the state $|\Psi_{out}\rangle$ is formed by the subsystems $R_1$, $C_2$, $K_1$ and $M_1$; $X\equiv \sigma^{(x)}$,   $Z\equiv \sigma^{(z)}$; we omit subscripts in notations $W^{(j)}$ for the brevity; operator $W^{(2)}$  is a novelty in comparison with earlier  algorithm  in \cite{ZBQKW_arXive2024}. Here the states $|\Psi^{(l)}\rangle$, $l=1,2$, are determined in Eq.(\ref{inst23}),  the state $|\Phi_0\rangle$ is determined in Eq.(\ref{Phi0}),
$W^{(0)}\equiv W^{(0)}_{C^{(1)}R^{(2)}}$ and $|\Phi_1\rangle$ are determined in Eqs.(\ref{W1m}) and (\ref{Phi11}),  $W^{(1)} \equiv W^{(1)}_{C^{(1)}}$ and $|\Phi_2\rangle$ are determined in Eqs.(\ref{W2m}) and (\ref{Phi32}), $W^{(2)}\equiv W^{(2)}_{M^{(1)}M^{(2)}K^{(1)}K^{(2)}}$ and $|\Phi_3\rangle$ are determined in Eqs.(\ref{W1}) and (\ref{Phi2}) , $W^{(3)}\equiv W^{(3)}_{C^{(1)}R^{(2)}M^{(2)}K^{(2)}B}$ and $|\Phi_4\rangle$ are determined in Eqs. (\ref{W3m}) and (\ref{Phi4}), $|\Phi_5\rangle$ and $|\Psi_{out}\rangle$ are determined in Eqs.(\ref{Phi6}) and (\ref{PsiOut}).  (b) The notation for the multi-qubit C-NOT. }
\label{Fig:Ap1}
\end{figure*}

\subsection{Manipulations with input matrices}
\label{Section:manipulations}
In  this section we consider several manipulations with the input  matrices $A^{(1)}$ and $A^{(2)}$. Instead of multiplication $A^{(1)}A^{(2)}$ we can perform
multiplications $(A^{(1)})^\dagger A^{(2)}$, $A^{(1)}(A^{(2)})^\dagger$, $(A^{(1)})^\dagger (A^{(2)})^\dagger$, $(A^{(1)}A^{(2)})^\dagger = (A^{(2)})^\dagger (A^{(1)})^\dagger$ and  change the order of the multipliers.
With this purpose we introduce the following operators.

To organize the Hermitian conjugate of  $A^{(i)}$, $i=1,2$, we use  operator $W$,   Eq.(\ref{SWAPM}), introduced in Sec.\ref{Section:HC} and rewrite it using the new notations:
\begin{eqnarray}\label{Q12}
Q^{(i)}_{ R^{(i)}C^{(i)}M^{(i)}}=  SWAP_{R^{(i)}C^{(i)}}  \sigma^{(z)}_{M^{(i)}} , \;\;i=1,2.
\end{eqnarray}
To change the order  of the matrices  in the multiplication algorithm we apply the operator
\begin{eqnarray}\label{Q3}
Q^{(3)}_{R^{(1)}C^{(1)}R^{(2)}C^{(2)}M^{(1)}M^{(2)}K^{(1)}K^{(2)}} =  SWAP_{R^{(1)}R^{(2)}} SWAP_{C^{(1)}C^{(2)}} SWAP_{M^{(1)}M^{(2)}}SWAP_{K^{(1)}K^{(2)}} .
\end{eqnarray}
Here, the multi-qubit $SWAP$-operators
$ SWAP_{R^{(1)}R^{(2)}}$ and $ SWAP_{C^{(1)}C^{(2)}}$ establish exchange between the appropriate qubits of two selected subsystems.
The circuit implementing  these operators is shown in Fig.\ref{Fig:manipulations}.
For instance, to calculate the product $A^{(1)}(A^{(2)})^\dagger$ we apply the operator $Q^{(2)}_{R^{(2)}C^{(2)}M^{(2)}}$ to the
state $|\Phi_0\rangle$, Eq.(\ref{Phi0}).
To calculate $(A^{(1)}A^{(2)})^\dagger \equiv (A^{(2)})^\dagger (A^{(1)})^\dagger$, we apply  the  operators
$Q^{(3)}_{R^{(1)}C^{(1)}R^{(2)}C^{(2)}M^{(1)}M^{(2)}K^{(1)}K^{(2)}}$, $Q^{(2)}_{R^{(2)}C^{(2)}M^{(2)}}$, $Q^{(1)}_{R^{(1)}C^{(1)}M^{(1)}}$  to the state $|\Phi_0\rangle$.
 Applying the   operators $Q^{(i)}$, $i=1,2,3$, can be controlled by additional qubits $q_{i}$ of the subsystem $q$  (see Fig.\ref{Fig:manipulations}), i.e.,  the following controlled SWAP
 operators (C-SWAP)  can be introduced:
\begin{eqnarray}\label{QS}
\hat Q^{(i)}_{S^{(i)}} = |1\rangle_{q_{i}}\,_{q_{i}}\langle 1| \otimes Q^{(i)}_{S^{(i)}} +  |0\rangle_{q_{i}}\,_{q_{i}}\langle 0| \otimes I_{S^{(i)}},\;\; i=1,2,3,
\end{eqnarray}
where $S^{(i)}= R^{(i)}C^{(i)}M^{(i)}$, $i=1,2$, $S^{(3)}=R^{(1)}C^{(1)}R^{(2)}C^{(2)}M^{(1)}M^{(2)}K^{(1)}K^{(2)}$ and $I_{S^{(i)}}$ is the
identity operator applied to $S^{(i)}$.
The state $|\varphi\rangle_q = |n_1\rangle_{q_{1}}  |n_2\rangle_{q_{2}}  |n_3\rangle_{q_{3}}$, $n_i=0,1$ ($i=1,2,3$), of the
controlling subsystem $q$ is  not changed after running the  algorithm. After applying the operators $\hat Q^{(j)}_{S^{(j)}}$, $j=1,2,3$, to $|\Phi_0\rangle|\varphi\rangle_q$, we obtain
\begin{eqnarray}\label{Phiq}
|\tilde \Phi_0\rangle =  \hat Q^{(3)}_{S^{(3)}} \hat Q^{(2)}_{S^{(2)}}    \hat Q^{(1)}_{S^{(1)}}
|\Phi_0\rangle|\varphi\rangle_q.
\end{eqnarray}
Of course calculating $(A^{(1)}A^{(2)})^\dagger$ can be done by performing the Hermitian conjugation after multiplication
$A^{(1)}A^{(2)}$, this is another option.

\begin{figure}[ht]
  \centerline{ \includegraphics[width=0.9\textwidth]{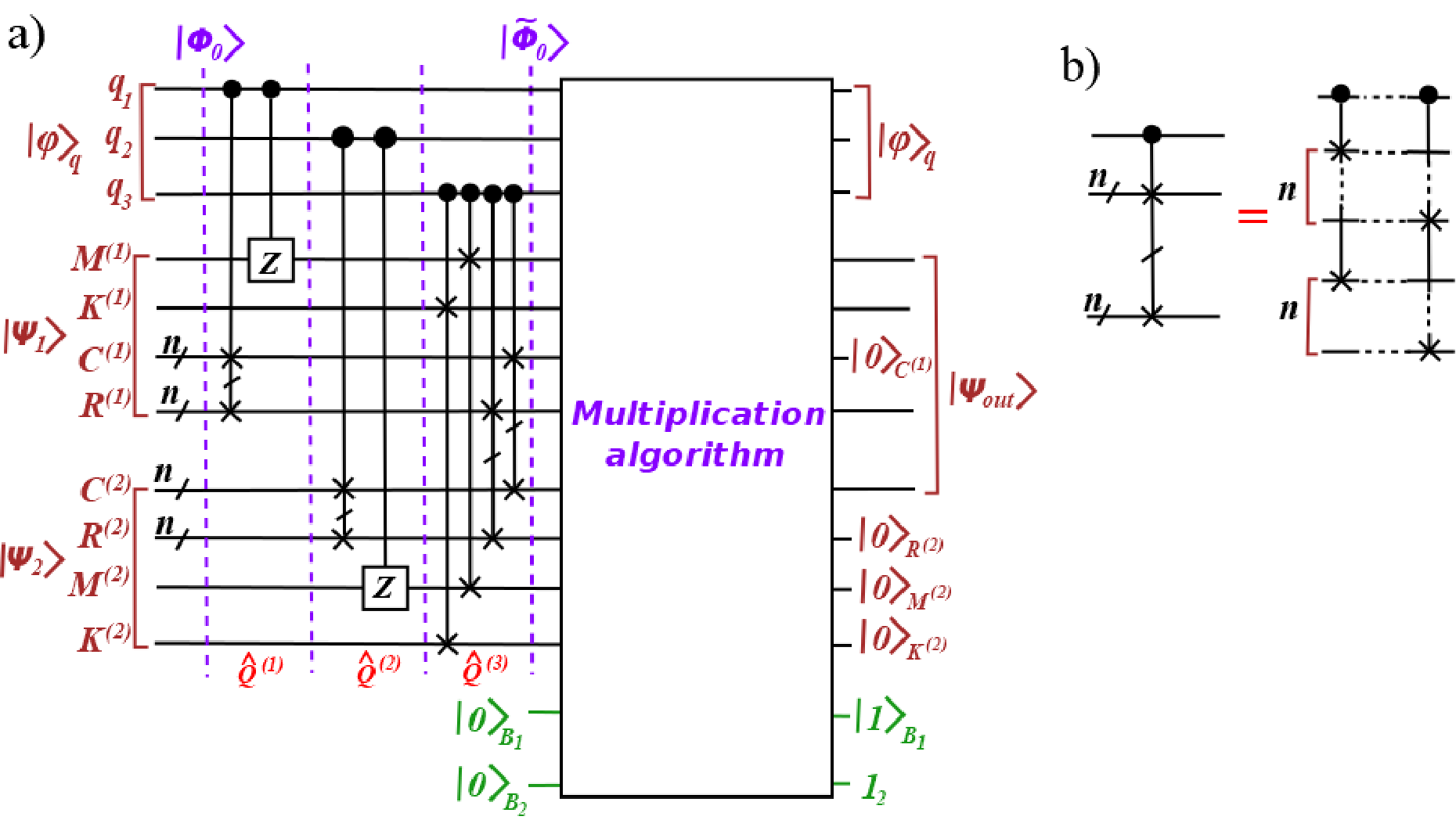}}
    \caption{(a) Controlled manipulations with input matrices $A^{(1)}$ and $A^{(2)}$. Here $|\varphi\rangle_q$ is determined above Eq. (\ref{Phiq}), the states $|\Psi^{(l)}\rangle$, $l=1,2$, are determined in Eq.(\ref{inst23}), the operators
     $\hat Q^{(j)}\equiv \hat Q^{(j)}_{S^{(j)}}$, $j=1,2,3$,  are determined in Eq. (\ref{QS}), the states
     $|\Phi_0\rangle$  and  $|\tilde \Phi_0\rangle$   are determined in Eqs.(\ref{Phi0}) and (\ref{Phiq}). The box ``Multiplication algorithm'' is represented by the circuit in Fig.\ref{Fig:Ap1}.   (b) Notation for the multi-qubit C-SWAP-operator; swaps are performed between the  $j$th qubits of each  subsystem, $j=1,\dots,n$.}
\label{Fig:manipulations}
\end{figure}

\subsection{Measuring normalization $G$ and discussion}
\label{Section:G}
Although  QCM allows to resolve the problem of exponential decrease of the success probability with the number of qubits encoding considered matrices, we lose the information about normalization $G$ in Eq.(\ref{PsiOut}). However, this normalization might be required in obtaining the final result. Below we explore  the method of probabilistic measuring the normalization $G$.

Let $S_1=|b^{(1)} b^{(2)}|^2$, this value is known due to the normalization in matrix encoding, see Eq.(\ref{constr2}).
If,  along with $S_1$, we know $S_1/G^2=\tilde S_1$, then $G=\sqrt{S_1/\tilde S_1}$. To find  $\tilde S_1$, we refer to  the state $|\Psi_{out}\rangle$, Eq.(\ref{PsiOut}), and measure the state of the subsystem $K_1$ with the output { $|0\rangle_{K_1}$}.  It is easy  to understand  that the probability of this output equals  $\tilde S_1$, and can be found in result of multiple runs of the algorithm. It is important that the problem of small success probability is resolved on passing from $|\Phi_4\rangle$, Eq.(\ref{Phi4}), to $|\Phi_5\rangle$, Eq.(\ref{Phi6}), and does not appear in the above measurement, {    where the success probability $\tilde S_1$  does not depend on  $n$}.

We have to note that the  last  measurement destroys the part of superposition quantum state in which the  result of  matrix multiplication is stored. Therefore, to get the full information, i.e., the normalization $G$ and matrix product, we have to use the set of runs to probabilistically determine $G$ and then run the algorithm one more time to end up with the state $|\Phi_5\rangle$ that includes  $|\Psi_{out}\rangle$ encoding the product $A^{(1)} A^{(2)}$. However, if $G$ is not needed, then the single run is enough.

We emphasize that both introduced extensions do not decline the parameters of the multiplication circuit such as depth and space.  Fig.\ref{Fig:Ap1} shows that the  depth is determined  by the operator $W^{(3)}_{C^{(1)} R^{(2)} M^{(2)}K^{(2)}B\tilde B} $, while the  manipulations with the input matrices  in Sec.\ref{Section:manipulations}, see  Fig.\ref{Fig:manipulations},  increase this depth only by insignificant constant factor. Thus, the estimation of the depth remains $O(n)$ (where $N=2^n$) similar to the estimation in \cite{ZBQKW_arXive2024}.
Regarding the space, the circuit in Fig.\ref{Fig:Ap1} includes only five   additional one-qubit subsystems ($K^{(i)}$, $M^{(i)}$, $i=1,2$ , and $B_2$) in comparison with the circuit in \cite{ZBQKW_arXive2024} with one of them (the ancilla qubit $B_2$) is needed for organization of   QCM. Therefore, the space estimation remains $O(n)$ as well. {    Of course, multiple running is required  in the case of implementing the usual ancilla-measurement \cite{ZBQKW_arXive2024} . Then  the depth of the whole algorithm is bigger than the depth of the circuit by $O(  2^{n+1}/G^2)$ times and equals $O(  2^{n+1}  n/G^2) \sim O(2^n n )$.}


\section{Conclusion}
\label{Section:conclusions}

{   In this paper we study  some new aspects of quantum algorithms with their application to  the   matrix-manipulation algorithms  proposed  in \cite{ZQKW_2024,ZBQKW_arXive2024}} and  based on the encoding the matrix elements  into the pure superposition   state of a quantum system.  { However, applicability of introduced novelties spreads much beyond of proposed examples.}

The most important { issue is the
replacement of}  the ordinary one-qubit ancilla-state  measurement (needed to remove the garbage acquired during evaluation of the quantum algorithm) with the measurement of the ancilla-state controlled by the state of another one-qubit ancilla (QCM).
This modification  allows to resolve the problem of small access probability to the desired state of the ancilla {(post-selection problem)}.
Remember, that the problem of small success probability  appears in the case of ordinary measurement of the ancilla state  in {  \cite{HHL,HHL1,HHL2,HHL3,HHL4,HHL5,HHL6,HHL7,ZQKW_2024,ZBQKW_arXive2024}.}
Thus,  QCM significantly increases the efficiency of the algorithms developed in those papers. At that,  we lose the  information on the normalization constant for the output state because this constant is expressed  in terms of   the access probability to the desired ancilla state  and this probability is not measured in the modified algorithm. However, { the  normalization  constant} can be revealed as shown in the matrix multiplication algorithm, Sec.\ref{Section:G}.
  We also have to note that the method of practical realization of  QCM is not developed  yet,  but its  realization would be one more justification of reality of superposition quantum states.

We also introduce two extensions { for  the quantum data encoding and apply them to the matrix-multiplication algorithm proposed in \cite{ZBQKW_arXive2024}.} The first extension    consists in  separating  the real and imaginary parts of the complex  matrix under consideration  by encoding these two parts into {  two orthogonal subspaces of the quantum    state-space.} Such extension allows constructing the  effective subroutine for operations typical for complex numbers, which  are complex and Hermitian conjugations.  Of course, this  separation of real and imaginary parts requires proper  modifications of the  quantum algorithms, { in particular, quantum algorithms} for  matrix addition, multiplication, determinant calculation, matrix inversion and solving linear systems that are described in \cite{ZQKW_2024,ZBQKW_arXive2024}.  In our paper, we present such modification for the multiplication algorithm and demonstrate manipulations with the input matrices using Hermitian conjugation  subroutine.

Our second extension weakens 
the normalization constraint, imposed on the elements of the  input matrices, which follows from  the normalization condition for a superposition pure quantum state. This can be done by  including an additional term  in the superposition state encoding  a particular matrix which  leads to   replacing  the  normalization equality, Eq.(\ref{constr0}), with inequality (\ref{constr1})  that imposes only the upper bound on  the sum of the  squared { modulus} of the matrix elements. { We emphasize that the applicability of  the two last extensions is not restricted to the matrix encoding. Any data-set  can be used instead of  matrices.}

Of course, each of  the above  modifications requires involving an extra qubit. Therefore, along with four subsystems $R^{(i)}$ and $C^{(i)}$,
$i=1,2$, enumerating rows and columns of the matrices  $A^{(i)}$, $i=1,2$,  the $i$th matrix is associated with  the one-qubit subsystem $M^{(i)}$, responsible for separate encoding  of the real and imaginary parts  of complex matrix elements, and the one-qubit  subsystem $K^{(i)}$ allowing to introduce an extra term into the superposition state encoding the matrix $A^{(i)}$, namely the presence of this extra term  weakens the normalization constraint. Thus, only four additional qubits appear in the  modified multiplication algorithm. In addition, involving  QCM requires an additional qubit for the ancilla $B$ which becomes the two-qubit ancilla.

We emphasize that all the above  modifications do not significantly  decline the depth and space of the {  circuit for} multiplication  algorithm. Estimations for both of them   remain $O(n)$, which is similar to the appropriate characteristics  of the circuit for the  multiplication algorithm in \cite{ZBQKW_arXive2024}.
However, the depth of the whole multiplication algorithm in \cite{ZBQKW_arXive2024}  is much bigger {because usual measurement is used therein which requires $O(2^n)$-times running of the algorithm to access the desired ancilla-state. }

Proposed  modifications can be included  in other matrix-manipulation  algorithms  considered in Refs.\cite{ZQKW_2024,ZBQKW_arXive2024}, { as well as in any { measurement-based algorithm \cite{HHL,WWLN,ZDB,ZBD,Wei,WeiBook} and algorithm based on data encoding \cite{ZZRF,BWPRWL,A,RML,WBL,SSP,Wang2}}}.

{\it Acknowledgments.} This project is supported by the National Natural Science Foundation of China (Grants No. 12031004,
No. 12271474 and No. 61877054). The work was also partially funded by a state task of Russian Fundamental Investigations
(State Registration No. 124013000760-0).

{
\section{Appendix: multiplication of two $4\times 4$ matrices}
\label{Section:Appendix}
Here we consider the multiplication of two $4\times 4$ matrices $(A^{(1)})^\dagger$ and $A^{(2)}$ with
\begin{eqnarray}
A^{(l)} =\left(
\begin{array}{cccc}
a^{(l)}_{00}&a^{(l)}_{01}&a^{(l)}_{02}&a^{(l)}_{03}\cr
a^{(l)}_{10}&a^{(l)}_{11}&a^{(l)}_{12}&a^{(l)}_{13}\cr
a^{(l)}_{20}&a^{(l)}_{21}&a^{(l)}_{22}&a^{(l)}_{23}\cr
a^{(l)}_{30}&a^{(l)}_{31}&a^{(l)}_{32}&a^{(l)}_{33}
\end{array}
\right),\;\;   a^{(l)}_{jk} = a^{(l)}_{jk0} + i a^{(l)}_{jk1}.
\end{eqnarray}
and use two  scalars $b^{(l)}=b^{(l)}_{0} + i b^{(l)}_{1}$, $l=1,2$, for the normalization purpose. In this case $n=2$, therefore  we use four $2$-qubit   subsystems   $R^{(l)}$ and $C^{(l)}$  and   four one-qubit subsystems $M^{(l)}$ and $K^{(l)}$,
 $l=1,2$.
Eq. (\ref{inst23}) yields
\begin{eqnarray}\label{inst23ex}
|\Psi^{(l)}\rangle &=&( b^{(l)}_{0}|0\rangle_{M^{(l)}} +b^{(l)}_{1}|1\rangle_{M^{(l)}})  |0\rangle_{R^{(l)}} |0\rangle_{C^{(l)}}|0\rangle_{K^{(l)}}
+ \\\nonumber
&&\sum_{j,k=0}^{3}  (a^{(l)}_{jk0}|0\rangle_{M^{(l)}} +a^{(l)}_{jk1}|1\rangle_{M^{(l)}})   |j\rangle_{R^{(l)}} |k\rangle_{C^{(l)}}|1\rangle_{K^{(l)}},\\\label{norm2ex}
&&|b^{(l)}|^2+\sum_{j,k=0}^3 |a^{(l)}_{jk}|^2=1,\;\;l=1,2.
\end{eqnarray}
The initial state $|\Phi_0\rangle$ given in  Eq.(\ref{Phi0}) reads
\begin{eqnarray}\label{Phi0ex}
&&|\Phi_0\rangle =  |\Psi_2\rangle \otimes |\Psi_1\rangle=
\\\nonumber &&
\sum_{m_1,m_2=0}^1
b^{(1)}_{m_1} b^{(2)}_{m_2} |m_1\rangle_{M^{(1)}}|m_2\rangle_{M^{(2)}}  |0\rangle_{R^{(1)}} |0\rangle_{C^{(1)}} |0\rangle_{R^{(2)}} |0\rangle_{C^{(2)}} |0\rangle_{K^{(1)}}|0\rangle_{K^{(2)}}+\\\nonumber
&&
   \sum_{{j_1,k_1,}\atop{j_2,k_2=0}}^{3}\sum_{m_1,m_2=0}^1
\left(a^{(1)}_{j_1k_1m_1}a^{(2)}_{j_2k_2m_2}|m_1\rangle_{M^{(1)}}|m_2\rangle_{M^{(2)}} \right)\\\nonumber
&&\times |j_1\rangle_{R_1} |k_1\rangle_{C^{(1)}} |j_2\rangle_{R^{(2)}} |k_2\rangle_{C^{(2)}}|1\rangle_{K^{(1)}}|1\rangle_{K^{(2)}}+
|g_0\rangle.
\end{eqnarray}
To calculate the Hermitian conjugate of $A^{(1)}$ we apply the operator $Q^{(1)}_{R^{(1)}C^{(1)}M^{(1)}}$, defined in Eq.(\ref{Q12}), to $|\Phi_0\rangle$ and obtain
\begin{eqnarray}\label{tPhi0ex}
&&|\tilde \Phi_0\rangle =
\\\nonumber &&
\sum_{m_1,m_2=0}^1 (-1)^{m_1}
b^{(1)}_{m_1} b^{(2)}_{m_2} |m_1\rangle_{M^{(1)}}|m_2\rangle_{M^{(2)}}  |0\rangle_{R^{(1)}} |0\rangle_{C^{(1)}} |0\rangle_{R^{(2)}} |0\rangle_{C^{(2)}} |0\rangle_{K^{(1)}}|0\rangle_{K^{(2)}}+\\\nonumber
&&
   \sum_{{j_1,k_1,}\atop{j_2,k_2=0}}^{3}\sum_{m_1,m_2=0}^1
\left( (-1)^{m_1}a^{(1)}_{k_1j_1m_1}a^{(2)}_{j_2k_2m_2}|m_1\rangle_{M^{(1)}}|m_2\rangle_{M^{(2)}} \right)\\\nonumber
&&\times |j_1\rangle_{R_1} |k_1\rangle_{C^{(1)}} |j_2\rangle_{R^{(2)}} |k_2\rangle_{C^{(2)}}|1\rangle_{K^{(1)}}|1\rangle_{K^{(2)}}+
|g_0\rangle.
\end{eqnarray}
  For the states  $|\Phi_j\rangle $, $j=1,\dots,4$, given in, respectively,   Eqs.(\ref{Phi11}), (\ref{Phi32}), (\ref{Phi2}) and  (\ref{Phi4}), we have
\begin{eqnarray}\label{Phi11ex}
&&
|\Phi_1\rangle =\\\nonumber
&&
\sum_{m_1,m_2=0}^1
(-1)^{m_1} b^{(1)}_{m_1} b^{(2)}_{m_2} |m_1\rangle_{M^{(1)}}|m_2\rangle_{M^{(2)}}  |0\rangle_{R^{(1)}} |0\rangle_{C^{(1)}} |0\rangle_{R^{(2)}} |0\rangle_{C^{(2)}}  |0\rangle_{K^{(1)}}|0\rangle_{K^{(2)}}+\\\nonumber
&&
\sum_{{j_1,j,}\atop{k_2=0}}^{3}\sum_{m_1,m_2=0}^1
\left((-1)^{m_1} a^{(1)}_{jj_1m_1}a^{(2)}_{jk_2m_2}|m_1\rangle_{M^{(1)}}|m_2\rangle_{M^{(2)}} \right)\\\nonumber
&&\times
  |j_1\rangle_{R^{(1)}} |j\rangle_{C^{(1)}} |0\rangle_{R^{(2)}} |k_2\rangle_{C^{(2)}}
|1\rangle_{K^{(1)}}|1\rangle_{K^{(2)}}+|g_1\rangle. 
\end{eqnarray}
\begin{eqnarray}\label{APhi3}
&&
|\Phi_2\rangle
 =\\\nonumber
&&\frac{1}{2}
\sum_{m_1,m_2=0}^1(-1)^{m_1}
b^{(1)}_{m_1} b^{(2)}_{m_2} |m_1\rangle_{M^{(1)}}|m_2\rangle_{M^{(2)}}  |0\rangle_{R^{(1)}} |0\rangle_{C^{(1)}} |0\rangle_{R^{(2)}} |0\rangle_{C^{(2)}}  |0\rangle_{K^{(1)}}|0\rangle_{K^{(2)}}\\\nonumber
&&+\frac{1}{2}
\sum_{j_1,k_2=0}^{3}  \sum_{ j=0}^{3}  \sum_{m_1,m_2=0}^1
\left((-1)^{m_1} a^{(1)}_{jj_1m_1}a^{(2)}_{jk_2m_2}|m_1\rangle_{M^{(1)}}|m_2\rangle_{M^{(2)}} \right)\\\nonumber
&&\times
  |j_1\rangle_{R^{(1)}} |0\rangle_{C^{(1)}} |0\rangle_{R^{(2)}} |k_2\rangle_{C^{(2)}}
|1\rangle_{K^{(1)}}|1\rangle_{K^{(2)}}+
|g_2\rangle. 
\end{eqnarray}
\begin{eqnarray}\label{Phi2ex}
&&
|\Phi_3\rangle
 =\\\nonumber
&&\frac{1}{2^{3/2}}\Big(
(b^{(1)}_{0} b^{(2)}_{0}+b^{(1)}_{1} b^{(2)}_{1})|0\rangle_{M^{(1)}} +
(b^{(1)}_{0} b^{(2)}_{1}-b^{(1)}_{1} b^{(2)}_{0})|1\rangle_{M^{(1)}}\Big) \\\nonumber
&&\times
 |0\rangle_{M^{(2)}} |0\rangle_{R^{(1)}} |0\rangle_{C^{(1)}} |0\rangle_{R^{(2)}} |0\rangle_{C^{(2)}}|0\rangle_{K^{(1)}}|0\rangle_{K^{(2)}} \\\nonumber
&&
+\frac{1}{2^{3/2}} \sum_{j_1,k_2=0}^{3} \sum_{j=0}^{3}   \left((a^{(1)}_{jj_10}a^{(2)}_{j k_20}+
a^{(1)}_{j j_1 1}a^{(2)}_{j k_2 1})|0\rangle_{M^{(1)}}
\right.\\\nonumber
&&\left.+
(a^{(1)}_{jj_10}a^{(2)}_{jk_21}-a^{(1)}_{jj_11}a^{(2)}_{jk_20})|1\rangle_{M^{(1)}} \right) \\\nonumber
&&\times
|0\rangle_{M^{(2)}}   |j_1\rangle_{R^{(1)}} |0\rangle_{C^{(1)}} |0\rangle_{R^{(2)}} |k_2\rangle_{C^{(2)}}|1\rangle_{K^{(1)}}|0\rangle_{K^{(2)}}  +
|g_3\rangle.
\end{eqnarray}
\begin{eqnarray}\label{Phi4ex}
&&
|\Phi_4\rangle =
\\\nonumber
&&
\frac{1}{2^{ 3/2}}\Big(
(b^{(1)}_{0} b^{(2)}_{0}+b^{(1)}_{1} b^{(2)}_{1})|0\rangle_{M^{(1)}} +
(b^{(1)}_{0} b^{(2)}_{1}-b^{(1)}_{1} b^{(2)}_{0})|1\rangle_{M^{(1)}}\Big) \\\nonumber
&&\times |0\rangle_{M^{(2)}}
 |0\rangle_{R^{(1)}} |0\rangle_{C^{(1)}} |0\rangle_{R^{(2)}} |0\rangle_{C^{(2)}}|0\rangle_{K^{(1)}}|0\rangle_{K^{(2)}} |1\rangle_{ B_1}  |1\rangle_{B_2} +\\\nonumber
&&
\frac{1}{2^{ 3/2}}\sum_{{j_1,k_2,}\atop{j=0}}^{3}
\Big((a^{(1)}_{jj_10}a^{(2)}_{jk_20}+a^{(1)}_{jj_11}a^{(2)}_{jk_21})|0\rangle_{M^{(1)}} +(a^{(1)}_{jj_10}a^{(2)}_{jk_21}-a^{(1)}_{jj_11}a^{(2)}_{jk_20})|1\rangle_{M^{(1)}} \Big)
\\\nonumber
&&\times
|0\rangle_{M^{(2)}}   |j_1\rangle_{R^{(1)}} |0\rangle_{C^{(1)}} |0\rangle_{R^{(2)}} |k_2\rangle_{C^{(2)}}|1\rangle_{K^{(1)}}|0\rangle_{K^{(2)}}\ |1\rangle_{B_1}|1\rangle_{B_2}+  |g_3\rangle 
 |0\rangle_{B_1} |0\rangle_{B_2}.
\end{eqnarray}
After removing the garbage via the measurement (either usual or quantumly controlled) we and up with the state
$|\Phi_5\rangle$, see Eq.(\ref{Phi6}):
\begin{eqnarray}\label{Phi6ex}
&&
|\Phi_5\rangle =
|\Psi_{out}\rangle\,|0\rangle_{M_2} |0\rangle_{C_1} |0\rangle_{R_2}|0\rangle_{K_2}|1\rangle_{B_1}  ,
\\\label{PsiOut2}
&&
|\Psi_{out}\rangle =\\\nonumber
&&
G^{-1}\left( (\hat b_0 |0\rangle_{M^{(1)}}    +\hat b_1|1\rangle_{M^{(1)}} )  |0\rangle_{R^{(1)}} |0\rangle_{C^{(2)}}  |0\rangle_{K^{(1)}} \right. \\\nonumber
&&\left.+
\sum_{j,k=0}^{N-1}(\hat a_{jk0}|0\rangle_{M^{(1)}} +\hat a_{jk1}|1\rangle_{M^{(1)}})| j\rangle_{R^{(1)}} |k\rangle_{C^{(2)}}|1\rangle_{K^{(1)}} \right),
\end{eqnarray}
where  $\displaystyle G=\left(|b^{(1)} b^{(2)}|^2+\sum_{j,k=0}^{3} \Big| \sum_{l=0}^{3} (a^{(1)}_{lj})^*a^{(2)}_{lk}\Big|^2 \right)^{1/2}$ is the normalization factor,
\begin{eqnarray}
&&
\hat b_0 = b^{(1)}_{0} b^{(2)}_{0}+b^{(1)}_{1} b^{(2)}_{1},\;\;
 \hat b_1=b^{(1)}_{0} b^{(2)}_{1}-b^{(1)}_{1} b^{(2)}_{0},\\\nonumber
&&
\hat a_{jk0} =  \sum_{l=0}^{3}
(a^{(1)}_{lj0}a^{(2)}_{lk0}+a^{(1)}_{lj1}a^{(2)}_{lk1}),\;\;
\hat a_{jk1} = \sum_{l=0}^{3}(a^{(1)}_{lj0}a^{(2)}_{lk1}-a^{(1)}_{lj1}a^{(2)}_{lk0}),
\end{eqnarray}
so that $\hat A=( A^{(1)})^\dagger A^{(2)} = \{\hat a_{jk}: j,k=0,\dots,3 \}$, $\hat a_{jk} = \hat a_{jk0} + i  \hat a_{jk1}$,
$\hat b= \hat b_0 + i \hat b_1$.
In particular, let
\begin{eqnarray}
&&
A^{(1)}=
\left(
\begin{array}{cccc}
 0.26 +0.04 i & 0.15 +0.24 i & 0.25 +0.18 i & 0.08\, +0.1 i
   \\
 0.03 +0.02 i & 0.18 +0.09 i & 0.24+0.1 i & 0.17 +0.1 i
   \\
 0.01 +0.25 i & 0.12+0.22 i & 0.01+0.21 i & 0.19 +0.23 i
   \\
 0.01 +0.17 i & 0.03+0.25 i & 0.03 +0.26 i & 0.22+0.15 i
   \\
\end{array}
\right),\\\nonumber
&&
A^{(2)}=\left(
\begin{array}{cccc}
 0.3+0.27 i & 0.04 +0.06 i & 0.02 +0.15 i & 0.14 +0.25 i
   \\
 0.17 i & 0 & 0.29+0.23 i & 0.08+0.04 i \\
 0.28+0.03 i & 0.04 +0.06 i & 0.24+0.03 i & 0.1+0.19 i
   \\
 0.24 +0.06 i & 0.05+0.12 i & 0.22+0.21 i & 0.22+0.18 i
   \\
\end{array}
\right),\\\nonumber
&&
b^{(1)}=0.317,\;\;\;
b^{(2)}=0.335.
\end{eqnarray}
Here $A^{(j)}$, $j=1,2$, are the exact random matrices and the parameters $b^{(j)}$, $j=1,2$,  are cut to satisfy the normalization Eq.(\ref{norm2ex}) up to the 3rd decimal.
Then, the algorithm yields
\begin{eqnarray}
\hat A=\left(
\begin{array}{cccc}
 0.1151, -0.0466 i & 0.0491 -0.0027 i & 0.0723 -0.0557 i &
   0.1309 +0.0003 i \\
 0.1875-0.1171 i & 0.0699-0.0111 i & 0.2064-0.0649 i &
   0.2044 -0.0449 i \\
 0.1725, -0.0648 i & 0.0665 -0.0094 i & 0.1945 -0.0409 i &
   0.1975\, -0.032 i \\
 0.1899 -0.061 i & 0.0596 +0.0219 i & 0.2213 -0.0162 i &
   0.1919 +0.0245 i \\
\end{array}
\right), \;\;\hat b = 0.106, \;\;G=0.669.
\end{eqnarray}
Here, the  parameter { $\hat b$ is  auxiliary}.
}

\end{document}